\documentclass[reprint,aps,pra]{revtex4-2}

\usepackage{amsmath}    % need for subequations
\usepackage{graphicx}   % need for figures
\usepackage{verbatim}   % useful for program listings
\usepackage{color}      % use if color is used in text
\usepackage{subfigure}  % use for side-by-side figures
\usepackage{hyperref}   % use for hypertext links, including those to external documents and URLs
\usepackage{units}
\usepackage{braket}
\usepackage[dvipsnames]{xcolor}
\usepackage{ulem}
\usepackage{siunitx}

\begin{document}

\title{Bichromatic state-dependent disordered potential for Anderson localization of ultracold atoms}


\author{Baptiste Lecoutre}\affiliation{Laboratoire Charles Fabry, Institut d'Optique, CNRS, Universit\'e Paris-Saclay, 91127 Palaiseau Cedex, France}

\author{Yukun Guo}\affiliation{Laboratoire Charles Fabry, Institut d'Optique, CNRS, Universit\'e Paris-Saclay, 91127 Palaiseau Cedex, France}

\author{Xudong Yu}\affiliation{Laboratoire Charles Fabry, Institut d'Optique, CNRS, Universit\'e Paris-Saclay, 91127 Palaiseau Cedex, France}

\author{M. Niranjan}\affiliation{Laboratoire Charles Fabry, Institut d'Optique, CNRS, Universit\'e Paris-Saclay, 91127 Palaiseau Cedex, France}

\author{Musawwadah Mukhtar}\affiliation{Laboratoire Charles Fabry, Institut d'Optique, CNRS, Universit\'e Paris-Saclay, 91127 Palaiseau Cedex, France}

\author{Valentin V. Volchkov}\affiliation{Laboratoire Charles Fabry, Institut d'Optique, CNRS, Universit\'e Paris-Saclay, 91127 Palaiseau Cedex, France}

\author{Alain Aspect}\affiliation{Laboratoire Charles Fabry, Institut d'Optique, CNRS, Universit\'e Paris-Saclay, 91127 Palaiseau Cedex, France}

\author{Vincent Josse}\email{vincent.josse@universite-paris-saclay.fr}
\affiliation{Laboratoire Charles Fabry, Institut d'Optique, CNRS, Universit\'e Paris-Saclay, 91127 Palaiseau Cedex, France}

%
%%\author[2,3]{\fnm{Second} \sur{Author}}\email{iiauthor@gmail.com}
%%\equalcont{These authors contributed equally to this work.}
%\author[1]{\fnm{Third} \sur{Author}}\email{iiiauthor@gmail.com}
%%\equalcont{These authors contributed equally to this work.}

%\affilliation{Laboratoire Charles Fabry, Institut d'Optique, CNRS, Universit\'e Paris-Saclay, 91127 Palaiseau Cedex, France}
\date{\today}

\begin{abstract}
The ability to load ultracold atoms at a well-defined energy in a disordered potential is a crucial tool to study quantum transport, and in particular Anderson localization. In this paper, we present a new method for achieving that goal by rf transfer of atoms of an atomic Bose-Einstein condensate from a disorder insensitive state to a disorder sensitive state. It is based on a bichromatic laser speckle pattern, produced by two lasers whose frequencies are chosen so that their light-shifts cancel each other in the first state and add-up in the second state. Moreover, the spontaneous scattering rate in the disorder-sensitive state is low enough to allow for long observation times of quantum transport in that state. We theoretically and experimentally study the characteristics of the resulting potential.
\end{abstract}

%\keywords{keyword1, Keyword2, Keyword3, Keyword4}



\maketitle

%%%%%%%%%%%%%%%%%%%%%%%%%%%%%%%%%%%%%%%%%%%%%%%%%%%%%%%
%%%%%%% INTRODUCTION
%%%%%%%%%%%%%%%%%%%%%%%%%%%%%%%%%%%%%%%%%%%%%%%%%%%%%%%
\section{Introduction}
\label{sc:section1}
Ultra-cold atoms offer remarkable quantum simulators to experimentally study difficult condensed-matter problems~\cite{bloch2008many}. Quantum transport and Anderson localization have been directly observed by launching atoms in disordered potentials produced by far off-resonance laser speckle~\cite{billy2008direct,roati2008anderson,jendrzejewski2012coherent,muller2015revival,kondov2011three,jendrzejewski2012three,semeghini2015measurement}. Convincing results with quantitative comparison to calculations have been obtained on Anderson localization in 1D speckle disordered potentials \cite{sanchez2007anderson,billy2008direct,roati2008anderson}, and on direct signature of weak localization phenomena~\cite{cherroret2012cbs,jendrzejewski2012coherent,muller2015revival}. When it comes to 3D Anderson localization, several observations have been reported \cite{kondov2011three,jendrzejewski2012three,semeghini2015measurement}, but precise quantitative measurements are still lacking. The reason is that in experiments performed so far, atoms launched in the disorder have a large energy dispersion so that evaluating the mobility edge demands a deconvolution leading to large uncertainties.

It would thus be extremely interesting to have a method to launch atoms in the disorder at a precisely defined energy. By scanning the energy around the mobility edge, it would allow one to determine precisely the mobility edge. One might even  evaluate  critical exponents. Note that  there is no exact theory yielding the value of both quantities~\cite{abrahams1979scaling,evers2008transitions}, and it is thus highly desirable to compare the results of approximate theoretical treatments~\cite{kuhn2007,skipetrov2008,piraud2013,piraud2014} or numerical results~\cite{delande2014,pasek2017anderson,slevin2014} with experimental results. Among other experiments that would benefit from such an improved control of the energy of the atoms launched in the disorder, one can cite measurements of  spectral functions~\cite{volchkov2018measurement}, tests of the landscape theory of localization~\cite{filoche2012}, search for sophisticated signatures of localization~\cite{karpiuk2012,micklitz2015,ghosh2017,hainaut2018,martinez2021}, or even the study of 2D localization~\cite{white2020,orso2017}.

\begin{figure}
\centering
\includegraphics[width=0.3\textwidth]{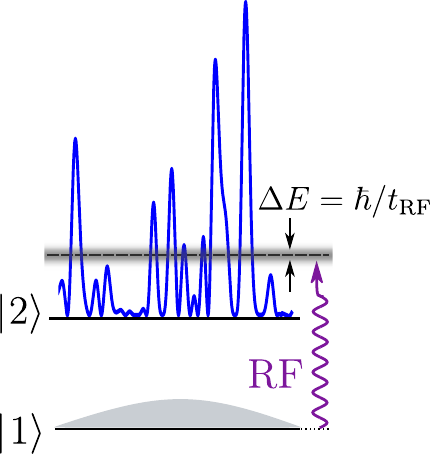}
\caption{Launching atoms at a well defined energy in a disordered potential. It consists of transferring atoms of a Bose-Einstein condensate from the discrete, disorder insensitive state $\lvert 1\rangle$ with a well defined energy,  to a state $\lvert 2 \rangle$ sensitive to disorder, thus belonging to a continuum. By tuning the rf transfer frequency $\omega_\mathrm{rf}$ one can select the mean energy of the atoms transferred in $\lvert 2 \rangle$, while adjusting the rf power and duration of the transfer allows one to control the spread in energy $\Delta E$ of the transferred atoms.
}
\label{fig:principe}
\end{figure}

Our strategy for launching atoms with a precisely defined energy (see Fig.~\ref{fig:principe}) 
consists in performing a rf transition from a state insensitive to disorder (state $\lvert 1\rangle$) to a state sensitive to disorder (state $\lvert 2\rangle$). The energy of the populated eigenstates in the disorder can be adjusted by the control of the frequency of the rf. 
In the disorder, the atom energy levels form a continuum. 
Since the initial state is discrete, a priori one has a one way transition characterized by a rate $\Gamma$ given by the Fermi Golden Rule, which also sets the minimum energy dispersion of the transferred atoms. For an interaction time $t_\mathrm{rf}$ shorter than $\Gamma^{-1}$, as in the experiment reported below, the energy dispersion of the arrival states, i.e. of the atoms transferred to the continuum, is Fourier-limited and given by $\Delta E = \hbar / t_\mathrm{rf}$.

\begin{figure}
\centering
\includegraphics[width=0.32\textwidth]{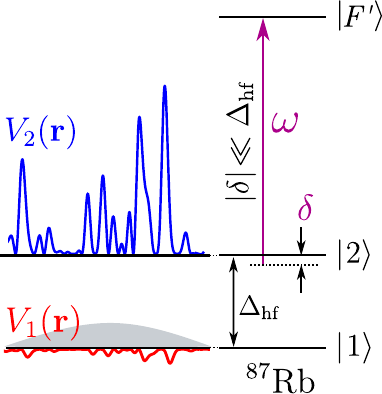}
\caption{State dependent disorder in a monochromatic speckle. The laser at frequency $\omega$ is closer to resonance for the $\lvert 2\rangle \rightarrow  \lvert F'\rangle$ transition than for the $\lvert 1\rangle \rightarrow  \lvert F'\rangle$ transition. The light shift induced in $\lvert 2\rangle$ is thus  larger than the light shift induced in $\lvert 1\rangle$, and for a speckle induced by that laser the resulting disorder is stronger in $\lvert 2\rangle$. On the figure, corresponding to a blue detuned laser, the disorder in $\lvert 2\rangle$  is repulsive, and is limited below. 
For a detuning $\delta$ of opposite sign (red detuned laser), the disorder would be attractive in $\lvert 2\rangle$, and limited above. The drawback of that method is the large spontaneous scattering of laser photons in state $\lvert 2\rangle$, resulting in a loss of coherence for the quantum transport of atoms in state $\lvert 2\rangle$.}
\label{fig:monochromatic}
\end{figure}

A first implementation of that scheme was demonstrated in the work of Volchkov et al.~\cite{volchkov2018measurement}, where we used a rf atom transfer between states with different sensitivities to a monochromatic laser speckle disorder. 
These different sensitivities stemmed from the very different detunings of the laser used to produce the disorder, for the initial state $\lvert 1\rangle$ and the final state $\lvert 2\rangle$ (see Fig.~\ref{fig:monochromatic}). The narrow Fourier-limited energy distribution we obtained --two orders of magnitude lower than in previous experiments \cite{pasek2017anderson,volchkov2018measurement}-- allowed us to make a quantitative study of the spectral function of the atoms in that disorder, exploring different regimes of quantum transport from the \textit{quantum} low disorder regime to the \textit{classical} strong disorder regime \cite{trappe2015semiclassical,prat2016semiclassical,pelletier2022spectral}. Striking differences between the cases of attractive (red detuned) and repulsive (blue detuned) disorder were observed and interpreted. The method of \cite{volchkov2018measurement} is, however, strongly limited by a serious problem. It relies on a laser tuned {\it between} the two resonances associated with the two atomic ground levels, and the detuning for the upper state $\lvert 2\rangle$ cannot be large enough to avoid resonant scattering of photons in that state. This entails a rapid destruction of the coherence of the spatial wave function describing the atomic motion and thus of Anderson localization.

\begin{figure}
\label{fig:illustration}
\centering
\includegraphics[width=0.28\textwidth]{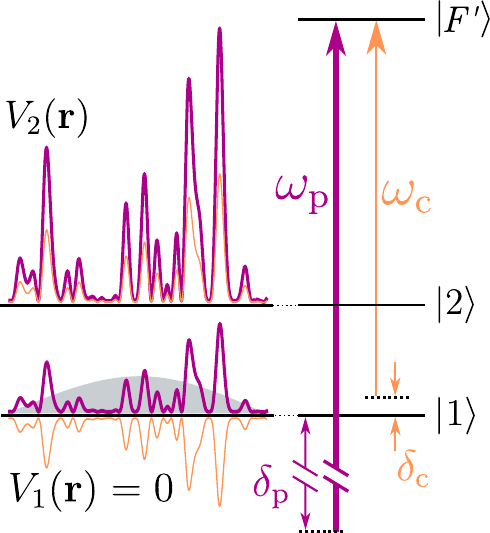}
\caption{State dependent disorder in a bichromatic speckle. The two lasers at frequencies $\omega_\mathrm{p}$ (principal) and $\omega_\mathrm{c}$ (compensation) create disordered potentials that add-up in $\lvert 2\rangle$ but cancel each other in $\lvert 1\rangle$. The detunings $\delta_\mathrm{p}$ and $\delta_\mathrm{c}$ are chosen so that the spontaneous scattering rate in $\lvert 2\rangle$ is small enough to have long coherence times in $\lvert 2\rangle$, i.e.,  on the order of one second or more. }
\label{fig:bichromatic}
\end{figure}

The method reported in the present paper overcomes this problem by the use of a bichromatic speckle potential. It consists of two speckles due to lasers of almost identical frequencies for which the potentials are of opposite signs for the initial state of the rf transfer, and of same signs for the final state of the rf transfer (Fig.~\ref{fig:bichromatic}). It yields a strong suppression of the sensitivity to the disorder in the initial state, together with a strong suppression of resonant scattering in the final -- disorder sensitive -- state. This scheme will allow one to operate with observation times around one second or more, which is required to study 3D localization phenomena~\cite{jendrzejewski2012three,semeghini2015measurement}.

This manuscript is organized as follows. In section \ref{sc:section2}, we recall  the most important properties of the state-dependent disordered potential based on a monochromatic laser speckle as used in Volchkov \textit{et al.}~\cite{volchkov2018measurement}, and discuss the limitations of that scheme. In section \ref{sc:section3}, we  describe the new method where we introduce a second laser to realize a bichromatic speckle. Firstly we study the influence of the small difference between the two laser frequencies creating the two speckles and estimate the fundamental potential decorrelation in the disorder sensitive state, an important result of this paper. Secondly we show how it is possible, by a suitable adjustment of the frequencies and intensities of the two lasers, to minimize the photon-scattering rate in the disorder sensitive state and to suppress the disorder experienced by the atoms in the initial state. In section~\ref{sc:experiment}, we present an experimental evaluation of the bichromatic speckle disorder scheme . The reported measurements support our analysis and are promising for future quantum transport experiments.

%%%%%%%%%%%%%%%%%%%%%%%%%%%%%%%%%%%%%%%%%%%%%%%%%%%%%%%
%%%%%%% SECTION 2: STATE-DEPENDANT POTENTIAL monochromatic
%%%%%%%%%%%%%%%%%%%%%%%%%%%%%%%%%%%%%%%%%%%%%%%%%%%%%%%

\section{State-dependent disordered  potential based on a monochromatic speckle} 
\label{sc:section2}

The creation of state dependent potential for alkali atoms has been widely investigated in the context of optical lattices using circularly polarized light tuned between the $D_1$ and $D_2$ lines, see e.g.~\cite{deutsch1998,mandel2003,gadway2010}. However such scheme is efficient only if the two considered state have different magnetic susceptibilities~\cite{grimm2000}. Since the study of 3D Anderson localization requires a magnetic levitation to suspend the atoms against gravity during their expansion in the disorder~\cite{kondov2011three,jendrzejewski2012three,semeghini2015measurement}, both disorder sensitive and insensitive states must have similar magnetic susceptibility.  Alternative methods to create state dependent disorder must then be developed, such as the monochromatic speckle state-dependent potential realized in Ref.~\cite{volchkov2018measurement}. This section recalls the main properties of that scheme and points out its limits.

\subsection{Monochromatic speckle potential}
\label{sc:section2_1}

\begin{figure}
\centering
\includegraphics[width=0.5\textwidth]{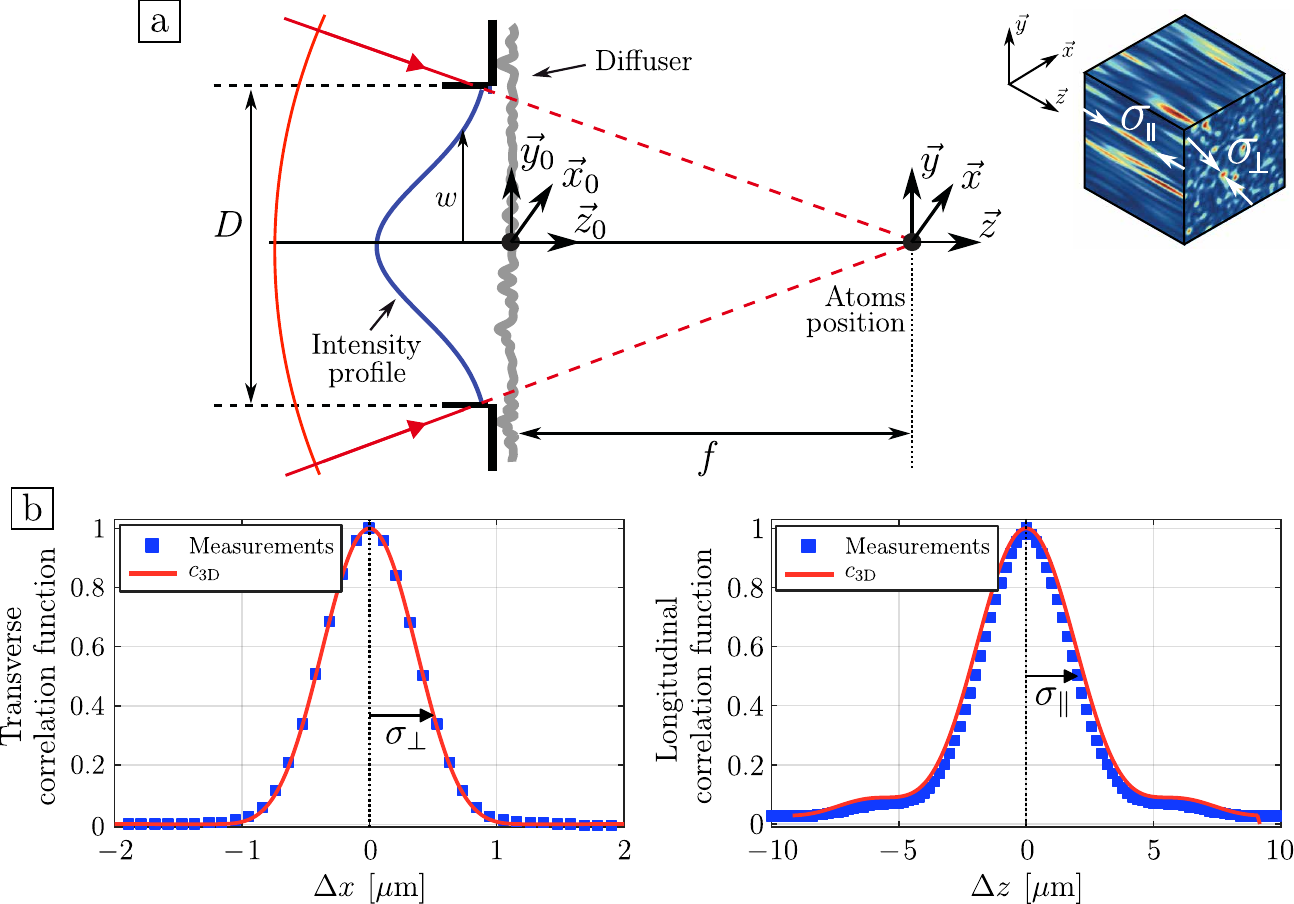}
\caption{a: Sketch of the generation of the speckle in an effective paraxial geometry. A laser beam of waist $w$ is focused using a lens of focal length $f=\SI{15.2 \pm 0.5}{\milli\meter}$. A rough plate of diameter $D$ scatters the laser light within an angle $\theta_\mathrm{diff}\approx 5^\circ$ which is fixed by the diffuser. The atomic cloud is centered on the optical axis in the Fourier plane $\lbrace x=0, y= 0, z= 0 \rbrace$. Inset: 3D view of a numerical realization of a speckle pattern. We define the size of the speckle grains by the half-widths $\sigma_\perp$ and $\sigma_\parallel$ of its autocorrelation function. b: Measured transverse $c_{3D}(\Delta\mathbf{r}_\perp,\Delta z = 0)$ and longitudinal $c_{3D}(\Delta\mathbf{r}_\perp = \mathbf{0}, \Delta z)$ autocorrelation functions of the speckle pattern, fitted by an effective paraxial theoretical model with effective parameters $D=\SI{17.8\pm 0.1}{\milli\meter}$ and $w=\SI{8\pm 1}{\milli\meter}$ \cite{richard2019elastic} (see text). This effective paraxial model successfully reproduces the features of the measured correlation functions.}
\label{fig:geometry_speckle}
\end{figure}

We consider the disordered potential  $V(\mathbf{r})$ experienced by an atom in a specific level in the presence of a monochromatic laser speckle pattern of intensity $I(\mathbf{r})$. For a non saturating  laser strongly detuned from the nearest relevant resonance, 
it scales as $I(\mathbf{r})/\delta$, where $\delta=\omega - \omega_0$ is the detuning of the laser with respect to the nearest resonant  transition involving the considered atomic level.  The speckle is generated by focusing an expanded laser beam onto the atoms located in the focal plane and diffracting it through a rough plate of random thickness, as illustrated in figure~\ref{fig:geometry_speckle}. The simplest model of such a speckle, as presented in \cite{goodman2007speckle}, results from the hypothesis that the rough plate imposes onto the laser beam a random phase $\phi(\mathbf{r}_\mathrm{0})$ with a  probability distribution constant over $2\pi$, and $\delta$-correlated --meaning that the spatial autocorrelation function of the transmission $t_\mathrm{diff}(\mathbf{r}_\mathrm{0})=\exp\{i\phi(\mathbf{r}_\mathrm{0})\}$ is a $\delta$ Dirac function. 

The complex amplitude $\mathcal{A}(\mathbf{r})$ of the field at each point of the fully developed speckle pattern --close to the focus of the laser-- can then be considered a sum of a very large number of independent random variables with the same statistical properties. According  to the central limit theorem, the complex amplitude is thus a Gaussian random process, whose properties allow one to calculate the statistical properties of the intensity $I(\mathbf{r})$, which is proportional to the squared modulus $\lvert\mathcal{A}(\mathbf{r})\rvert^2$ of the complex amplitude. Since  the complex amplitude at each point  has a 2-dimensional Gaussian probability distribution, the intensity $I(\mathbf{r})$ has an exponential probability distribution  $\mathcal{P}(I)=\overline{I}^{-1} \exp[-I/\overline{I}] \Theta (I/\overline{I})$ \cite{goodman2007speckle} with $\Theta$ the unit step function, and where the symbol $\overline{\:\cdots\:}$ stands for ensemble averaging, i.e., averaging over different realizations of  the rough plate. The standard deviation of the intensity fluctuations is then equal to the mean intensity value, i.e., $\sigma_I=\overline{I}$. A similar property $  \sigma_V =\lvert \overline{V} \rvert$ holds for the disordered  potential which can be positive or negative. It is experimentally controlled  by adjusting the laser power and detuning,  and can be varied over several orders of magnitude \cite{volchkov2018measurement}. Both $\overline{I}$ and $\overline{V}$ are independent of $\mathbf{r}$ because the process is spatially invariant and ergodic for this simple model of uncorrelated dephasing.

A key property of the fully developed speckle pattern is the size of the speckle grains, which is equal, within a factor, to the half-width of the normalized autocorrelation function defined around the focal point where the atoms are located
\begin{equation}
c_{3D}(\Delta\mathbf{r}_\perp, \Delta z) = \frac{\overline{\delta I(\mathbf{0},0) \, \delta I (\Delta\mathbf{r}_\perp, \Delta z)}}{\overline{\delta I^2}} 
\label{eq:c3D definition}
\end{equation}
Here $\delta I(\mathbf{r})=I(\mathbf{r})- \overline{I} $ are the intensity fluctuations, and we note $\Delta \mathbf{r}_\perp = \lbrace \Delta x, \Delta y \rbrace$ the transverse displacement in the Fourier plane. The function $c_{3D}$ is a fourth order moment of the complex amplitude $\mathcal{A}(\mathbf{r})$, and for the Gaussian process considered here, one can use the Wick's  theorem for classical moments of a Gaussian random process, to express any moment of $I(\mathbf{r})$ as a function of the second order moments of the amplitude \cite{goodman2007speckle}. The function $c_{3D}$ can then be expressed as a function of the autocorrelation function of the complex amplitude $\Gamma_{\mathcal{A}}(\Delta\mathbf{r}_\perp, \Delta z)$ (calculated in section \ref{A2} of Appendix A) as
\begin{eqnarray}
\label{eq:wick_c3D}
c_{3D}(\Delta\mathbf{r}_\perp, \Delta z) &=& \frac{\left\lvert \Gamma_{\mathcal{A}}(\Delta\mathbf{r}_\perp, \Delta z)\right\rvert^2}{\left\lvert \Gamma_\mathcal{A}(\mathbf{0},0)\right\rvert^2} \nonumber \\ 
&=&  \frac{\left\lvert\mathrm{FT} \left[I(\mathbf{r}_0) e^{-i \frac{\mathbf{r}_\mathrm{0}^2 k \Delta z}{2f^2}} \right]_{\frac{k\Delta\mathbf{r}_\perp}{f}} \right\rvert^2}{\left\lvert\displaystyle\int\mathrm{d}\mathbf{r}_\mathrm{0} \, I(\mathbf{r}_\mathrm{0}) \right\rvert^2} 
\end{eqnarray}
This correlation function is central to quantum transport studies as it directly translates into the spatial frequency distribution of the potential that governs the scattering properties of the atoms~\cite{sanchez2007anderson,kuhn2007coherent,shapiro2012cold,richard2019elastic,signoles2019ultracold}. A full calculation is presented in section \ref{A3} of Appendix A, and here we only discuss some of its most important features in the transverse and longitudinal directions.

Within the paraxial approximation, the transverse amplitude correlation function $\Gamma_{\mathcal{A}}(\Delta\mathbf{r}_\perp, \Delta z=0)$ is the Fourier transform of the intensity distribution $I(\mathbf{r}_\mathrm{0})$ just before the diffusing plate \cite{goodman2007speckle}. This can be interpreted as an example of the Van Cittert-Zernike theorem \cite{born2013principles}. The intensity fluctuations correlation function is thus nothing else than the squared modulus of the diffraction pattern corresponding to the intensity distribution at the diffusing plate in the diffuser
\begin{equation}
\label{eq:c3D_transverse}
c_{3D} (\Delta\mathbf{r}_\perp, \Delta z = 0) \propto \left\lvert \mathrm{FT} \left[ I(\mathbf{r}_\mathrm{0}) \right]_{\frac{k\Delta\mathbf{r}_\perp}{f}} \right\rvert^2 \, .
\end{equation}
For instance, in the ideal case of a Gaussian intensity profile of standard waist radius $w$ (following the usual convention for laser beams, $w$ is the radius at $\mathrm{e}^{-2}$), it is 
\begin{equation}
\label{eq:c3D_transverse_gauss}
c_{3D}(\Delta\mathbf{r}_\perp, \Delta z=0)= \exp\left \{-\frac{\Delta\mathbf{r}_\perp^2}{\sigma^2_\perp}    \right \}
\, ,
\end{equation}
with $\sigma_\perp=\lambda f /\pi w$. We call $\sigma_\perp$  the size of a speckle grain, for consistency with previous papers. Within the same paraxial hypothesis, the longitudinal autocorrelation function of the intensity fluctuations is, on the $z$ axis and close to the focusing point of the lens \cite{goodman2007speckle}
\begin{equation}
\label{eq:c3D_longi}
c_{3D}(\mathbf{0}, \Delta z)\propto
\left\lvert \int{\mathrm{d}\mathbf{r}_\mathrm{0} I(\mathbf{r}_\mathrm{0})\exp\left \{-i k \frac{\mathbf{r}_\mathrm{0}^2\Delta z}{2 f^2} \right  \}}\right\rvert^2 .
\end{equation}
In general, the evaluation of the integral in (\ref{eq:c3D_longi}) can be done numerically only. In the ideal case of a Gaussian intensity profile, it yields a Lorentzian profile of the longitudinal correlation function \cite{goodman2007speckle} and we define the size of a speckle grain by the half-width at half-maximum $\sigma_\parallel$.

A more realistic description of the situation considers a correlated diffuser with a spatial autocorrelation function of the phase factor -- more precisely of $\exp\{ i\phi(\mathbf{r}_\mathrm{0})\}$ -- of finite width $r_\mathrm{e}$ (see section~\ref{A1} of the appendix or Ref.~\cite{goodman2007speckle}). The speckle pattern has then a mean intensity profile that is no longer uniform, in contrast to the $\delta$-correlated model above. This profile is given by the Fourier transform of the autocorrelation function of the phase factor. For our diffuser, it has a Gaussian shape with an angular width of about $\theta_\mathrm{diff}\approx 5^\circ$ ($1/e^2$ radius) and yields a speckle pattern extending on a zone much wider ($\sim \SI{1.3}{\milli\meter}$) than the atomic cloud ($\sim \SI{15}{\micro\meter}$). The average intensity on the atoms is then almost constant
and the speckle can still be considered uniform over the atoms sample. Moreover, the correlation length of the phase factor of the rough plate is small compared to the width $w$ of the laser beam on the plate, so that the central limit theorem still applies to the speckle complex amplitude which is well represented by a Gaussian statistics. We can thus use the formulae (\ref{eq:c3D_transverse}) and (\ref{eq:c3D_longi}).

Another more realistic feature of our experiment is the fact that the phase distribution at the diffuser may not be strictly uniform over $2\pi$. In fact, as shown in \cite{lecoutre2020transport} and in section \ref{A1} of the Appendix A, a model with a Gaussian phase distribution of standard deviation $\sigma_\phi$ allows us to show that the formulae (\ref{eq:c3D_transverse}) and (\ref{eq:c3D_longi}) can still be used as long as $\sigma_\phi\gg 2\pi$, for a monochromatic speckle. We will see in section \ref{sc:section3} that a bichromatic speckle requires a more elaborated description, for which we will use the model of a plate with a Gaussian distribution of thickness.

The experimental determination of the correlation function (\ref{eq:c3D definition}) was done with an optical microscope~\cite{richard2015propagation} and the results are shown in Figure~\ref{fig:geometry_speckle}.b. Because of the large numerical aperture $\mathrm{NA}=0.55(2)$, the theoretical description requires a beyond-paraxial model~\cite{volchkov2018measurement}. In fact, the measured transverse and longitudinal correlation functions were found well reproduced by equations (\ref{eq:c3D_transverse}) and (\ref{eq:c3D_longi}), provided that we introduce a geometrical scaling factor~\cite{richard2019elastic,richard2019sm}. More precisely, we can match the measured correlation functions using formulaes (\ref{eq:c3D_transverse}) and (\ref{eq:c3D_longi}) with a geometrical factor of $0.875$, resulting in the effective numerical aperture $\mathrm{NA}_\mathrm{eff}=0.50$. In particular, the transverse profile of the three-dimensional correlation function is very well described by a gaussian function (\ref{eq:c3D_transverse_gauss}) of $1/e$ radius $\sigma_\perp=\SI{0.50\pm 0.01}{\micro\meter}$.
The longitudinal profile is also well reproduced by the result of a numerical evaluation of (\ref{eq:c3D_longi}) based on a truncated gaussian illumination~\footnote{The case of a pure gaussian intensity profile on the diffuser leads to a Lor?entzian longitudinal correlation function while it is described by a $\mathrm{sinc}^2$ function in the case of a circularly truncated homogeneous illumination. In both cases, we characterize its width by the HWHM.}
 \cite{richard2019elastic}. A longitudinal correlation length can still be defined by the half-width at half-maximum, yielding $\sigma_\parallel=\SI{2.05\pm 0.05}{\micro\meter}$ \cite{volchkov2018measurement}.
Altogether, this indicates that an effective paraxial approximation is well-suited to describe the spatial correlations of our speckle.

\subsection{Energy-resolved transfer scheme in a monochromatic state-dependant optical potential for $^{87}\mathrm{Rb}$}
\label{sc:section2_2}

In this subsection, we address the two problems that influence the energy spread of $^{87}\mathrm{Rb}$ atoms transfered into a disordered potential: the fluctuations of the energy difference between the magnetic levels due to magnetic field fluctuations, and the energy dispersion of the initial state.

\subsubsection{Suppression of the effect of the magnetic field fluctuations}
\label{sc:section2_2_1}
The atomic sample consists of a Bose-Einstein Condensate (BEC) prepared in the disorder insensitive state $\lvert 1\rangle$ with energy $E_1$ and then coupled by a rf field to the disorder sensitive state $\lvert 2\rangle$ with energy $E_2$. The final kinetic energy $E_f$ relevant to the transport is $E_f=E_i+\hbar \delta_\mathrm{rf}$ where $E_i$ is the initial kinetic energy, which is null for atoms initially at rest, and 
  $\delta_\mathrm{rf}=\omega_\mathrm{rf}-\Delta_\mathrm{hf}$ is  the detuning of the rf field with respect to the bare $\lvert 1\rangle\rightarrow \lvert 2\rangle$ transition. The energy $E_f$ can be chosen by tuning the rf frequency, allowing us to address  precise energy values  of the  atoms in the disorder-sensitive state. 

If the bare resonance frequency $\Delta_\mathrm{hf}$ between the two magnetic sublevels fluctuates because of magnetic field fluctuations, the energy $E_f$ will also fluctuate. In order to cancel these fluctuations of $E_f$, we  use the so-called ``clock states" $\lvert 1\rangle \equiv \lvert F=1, m_\mathrm{F} =-1\rangle$ and $\lvert 2\rangle \equiv \lvert F=2, m_\mathrm{F} = +1 \rangle$,  which are separated at zero field by a splitting of $\Delta_\mathrm{hf}/2\pi= \SI{6.835}{\giga\hertz}$, and we impose a bias magnetic field at the so-called \textit{magic}  value of $B_0^*=3.23\,\mathrm{G}$. The magnetic susceptibilities of states $\lvert 1\rangle$ and $\lvert 2\rangle$ of $^{87}\mathrm{Rb}$ are then identical, meaning that the energy separation between these states is insensitive to magnetic field fluctuations~\cite{lewandowski2002}. Moreover, this choice of the clock states is also crucial regarding the use of the magnetic levitation to study the propagation of the atoms in 3D.

The rf coupling consists in fact of a two-photon transition, involving a microwave and a rf field, to match the angular momentum difference $\Delta m_F=2$ \cite{volchkov2018measurement}. It can be considered a direct transition induced by a  field with an  effective frequency $\omega_\mathrm{rf}$ equal to the sum of the two frequencies of the two fields, and an effective Rabi frequency proportional to the product of the Rabi frequencies of the two fields.

\subsubsection{Suppression of the sensitivity of the initial state to the disordered potential}
\label{sc:section2_2_2}
In order to have a speckle acting strongly on the atomic state $\lvert 2\rangle$ but very little on the atomic state $\lvert 1\rangle$, we use a laser of frequency $\omega$ close to resonance for the transition $\lvert 2\rangle \rightarrow \lvert F'\rangle$ and far from resonance for the transition $\lvert 1\rangle \rightarrow \lvert F'\rangle$ (see Fig.~\ref{fig:monochromatic}). More precisely, since the natural linewidth of the transition ($\Gamma_\mathrm{Rb} /2\pi\simeq \SI{6.07}{\mega\hertz}$) is small compared to the splitting $\Delta_\mathrm{hf} /2\pi\simeq \SI{6.8}{\giga\hertz}$ \cite{steck2001rubidium}, it is possible to operate in the regime of $\Gamma_\mathrm{Rb} \ll\delta \ll \Delta_\mathrm{hf}$, with $\delta= \omega -\omega_{2,F'}$ the detuning of the speckle laser from resonance for state $\lvert 2 \rangle$. Then the detuning from resonance for state $\lvert 1\rangle$ is almost equal to $\Delta_\mathrm{hf}$, so that the average speckle potentials for level $\lvert 1\rangle$ and level $\lvert 2\rangle$ are in a ratio
\begin{equation}
\label{e222}
\frac{V_1}{V_2} \sim \frac{\delta}{\Delta_\mathrm{hf}} \ll 1 \,.
\end{equation}
With well chosen laser intensity $I$ and detuning $\delta$, one can then obtain an almost disorder-insensitive state $\lvert 1\rangle$, i.e., a disordered potential $V_1$  small compared to the chemical potential $\mu$, so that the screening effect in the Bose-Einstein condensate prepared in state $\lvert 1\rangle$  absorbs the residual potential $V_1$ and suppresses further its eventual perturbation~\cite{sanchez2006screening}. In contrast, state $\lvert 2\rangle$ is sensitive to the disorder, as expected. 
For instance, the choice of $\delta/2\pi \approx \pm \SI{80}{\mega\hertz}$ in Volchkov \textit{et al.} leads to a disordered potential ratio of $\lvert V_2 / V_1 \rvert \sim 100$ with $V_1 \leq \mu$.

This simple implementation of a state-dependent disordered potential allowed us to determine the spectral functions of ultra-cold atoms in a speckle potential at various energies. The results were found in  remarkable agreement with a numerical theoretical treatment in all regimes of disorder~\cite{volchkov2018measurement}. This result shows the interest of a transfer between a weakly disorder sensitive state and a strongly disorder sensitive state, in order to make  energy-resolved measurements on  transport of atoms in disorder.

However, equation~\eqref{e222} with the requirement  $V_1<\mu$ prevents one to have a large value of $\delta$, so that the spontaneous photon scattering rate in state $\lvert 2\rangle$,  which scales as $\Gamma\sim I / \delta^2$, is large enough to make it impossible to  study  quantum transport of the  atoms in state $\lvert 2\rangle$ for a long time. This is because  the spontaneous  scattering of a  photon by an atom in state $\lvert 2\rangle$  destroys the motional wave-function coherence, which is at the heart of quantum transport and Anderson localization. For instance, in the configuration of Volchkov \textit{et al.},  the photon scattering rate was as large as a few $10^3\, \mathrm{s}^{-1}$, forbidding  measurement times beyond a few milliseconds, see Table~\ref{tbl:mukhtar}. This inadequacy of the monochromatic speckle potential to allow for long-lasting energy-resolved quantum transport experiments can be overcome by the use of a bichromatic speckle potential, as explained in next section.

%%%%%%%%%%%%%%%%%%%%%%%%%%%%%%%%%%%%%%%%%%%%%%%%%%%%%%%
%%%%%%% SECTION 3: BICHROMATIC SPECKLE POTENTIAL PROPERTIES
%%%%%%%%%%%%%%%%%%%%%%%%%%%%%%%%%%%%%%%%%%%%%%%%%%%%%%%

\section{State-dependent disordered potential based on a bichromatic speckle}
\label{sc:section3}

In subsection~\ref{sc:bichromatic_principle}, we present the improved scheme to experimentally study quantum transport of atoms, based on a disordered potential created by two lasers at two different frequencies. In subsection \ref{sc:cancellation_potential}, we give a quantitative evaluation of the residual disorder due to a fundamental potential decorrelation resulting from the difference in frequencies between the two lasers. It is characterized by a normalized bichromatic correlation function, whose form is remarkable. 
It allows us to show that  the residual disorder in state $\lvert 1\rangle$ has the same order of magnitude as in the monochromatic scheme.
In subsection~\ref{sc:lifetime}, we evaluate the residual spontaneous scattering rate of laser photons in state $\lvert 2\rangle$, and we show that is reduced by several orders of magnitude, which is the main goal of the new scheme.

\subsection{Bichromatic speckle scheme}
\label{sc:bichromatic_principle}
As shown in figure~\ref{fig:bichromatic}, we use two lasers sufficiently detuned from resonance with state $\lvert 2\rangle$. The principal laser (purple in Fig.~\ref{fig:bichromatic}) is largely detuned and we fix its detuning $\delta_\mathrm{p}/2\pi = \SI{105}{\giga\hertz}$ with respect to the transition $\lvert 1\rangle\rightarrow \lvert F'\rangle$.
The potential on state $\lvert 1\rangle$ is suppressed by the use of a less-detuned compensation laser (orange in Figure~\ref{fig:bichromatic}), with an opposite detuning sign. We can add the two potentials without considering possible interferences between the lasers because the beat-note frequency ($\sim\SI{100}{\giga\hertz}$ in the implementations presented below) is  high enough  that it is averaged out by the inertia of the atoms and  has no effect on the atomic motion. 
In the linear regime, the total disordered potential for atoms in state $\lvert F,m_F\rangle$ can thus be written as
\begin{eqnarray}
V_{F,m_F}(\mathbf{r}) &=&-\frac{1}{2 \epsilon_0 c} \big( \; \mathrm{Re}[\alpha_{F,m_F}(\delta_\mathrm{p})] I_\mathrm{p}(\mathbf{r})  \nonumber \\
& & + \mathrm{Re}[\alpha_{F,m_F}(\delta_\mathrm{c}) ] I_\mathrm{c}(\mathbf{r})  \; \big) \; ,
\end{eqnarray}
where the complex atomic polarizability $\alpha_{F,m_F}(\delta)$ describes the atomic dipolar response of the internal state $\lvert F,m_F\rangle$ to an external electric field of detuning $\delta = \omega - \omega_{1,F'}$. The subscript $\mathrm{p}$ stands for principal and $\mathrm{c}$ stands for compensation.

Following the scheme sketched in Fig.~\ref{fig:bichromatic}, we want to adjust the two potentials such that they cancels each other in state $\lvert 1\rangle$, while summing up in state $\lvert 2\rangle$. We denote $V_\mathrm{R}$ the amplitude of the total laser speckle field in this disorder sensitive state, which is the relevant quantity when considering the study of quantum transport phenomena in such disordered potential.These two conditions write 
\begin{equation}
\left\lbrace \begin{array}{ll}
V_1(\mathbf{r}) = V_\mathrm{p,1}(\mathbf{r}) + V_\mathrm{c,1}(\mathbf{r}) \\
V_2(\mathbf{r}) = V_\mathrm{p,2}(\mathbf{r}) + V_\mathrm{c,2}(\mathbf{r})
\end{array}\right. \;  \mathrm{with} \;  \left\lbrace \begin{array}{ll}
\overline{V_1} =  0 \\
\overline{V_2} = V_\mathrm{R} 
\end{array} \right.
\label{eq:total_potential}
\end{equation}

Cancelling the total potential $V_1(\mathbf{r})$ on state $\lvert 1\rangle$ for all $\lbrace\mathbf{r}\rbrace$ requires the two speckle patterns to be as identical as possible. To do so, a first mandatory condition is to shine two identical laser modes on the diffuser (as shown in figure~\ref{fig:geometry_speckle} for a single laser beam). This is experimentally done by injecting the two lasers into the same optical fiber before shining them onto the diffuser. 

A second condition lies in the stability of the laser intensities. Power fluctuations of one of the beams of the order of a few percents can lead to strong potential fluctuations on state $\lvert 1\rangle$ -- up to a few percents of $V_\mathrm{R}$ -- which can be limiting, see table \ref{tbl:mukhtar}. Each of the lasers are therefore power-stabilized. 

However, strict equality of the two identical speckle patterns cannot be achieved with different wavelengths, because  the speckle patterns depend on diffraction and therefore on the laser wavelengths. The use of different laser wavelengths then leads to fundamental potential decorrelation preventing the complete cancelling of the disorder for state $\lvert 1\rangle$. That is the subject of the next section. 

\subsection{Fundamental potential decorrelation in the disorder insensitive state}
\label{sc:cancellation_potential}

Recalling that the detunings and intensities of the two lasers are adjusted such that $\overline{V_\mathrm{c,1}}=-\overline{V_\mathrm{p,1}}$, we want to characterize the residual fluctutations $\delta V_1(\mathbf{r})= \delta V_\mathrm{p,1}(\mathbf{r}) + \delta V_\mathrm{c,1}(\mathbf{r} )$ due to the frequency difference between the two lasers. We thus evaluate the variance of $V_1(\mathbf{r})$ at each point. For a speckle pattern, one has $\sigma_V=\overline{V}$, and the variance can be expressed as:
\begin{equation}
\sigma_{V_1}^2 (\mathbf{r})
= 2 \lvert \overline{V_\mathrm{p,1}}(\mathbf{r}) \, \overline{V_\mathrm{c,1}}(\mathbf{r})\rvert \left( 1- c_{2\lambda}(\mathbf{r}, \lambda_\mathrm{p}, \lambda_\mathrm{c}) \right) \mathrm{ ,}
\label{eq:variance_bichromatic}
\end{equation}
where the normalized bichromatic correlation function $c_{2\lambda}$ is defined as
\begin{equation}
c_{2\lambda}(\mathbf{r}, \lambda_\mathrm{p}, \lambda_\mathrm{c}) \equiv \frac{\overline{\delta V_\mathrm{p,1}(\mathbf{r}) \delta V_\mathrm{c,1}(\mathbf{r})}}{\overline{V_\mathrm{p,1}(\mathbf{r})}\:\overline{V_\mathrm{c,1}(\mathbf{r})}} \,.
\label{eq:bichromatic correlation}
\end{equation}
The correlation function (\ref{eq:bichromatic correlation}) quantifies the correlation of the two speckle fields at position $\mathbf{r}$ in space as a function of their wavelengths. In the case of totally decorrelated speckle fields ($c_{2\lambda}=0$), the variance of the total potential is consistently given by the sum of the variances of each field. In contrast, for exactly identical speckle patterns, the variance of the total potential would be null. Studying the amplitude of the residual potential then comes down to investigating the behavior of the normalized bichromatic correlation function $c_{2\lambda}(\mathbf{r}, \lambda_\mathrm{p}, \lambda_\mathrm{c})$.

\begin{figure}
\centering
\includegraphics[width=0.4\textwidth]{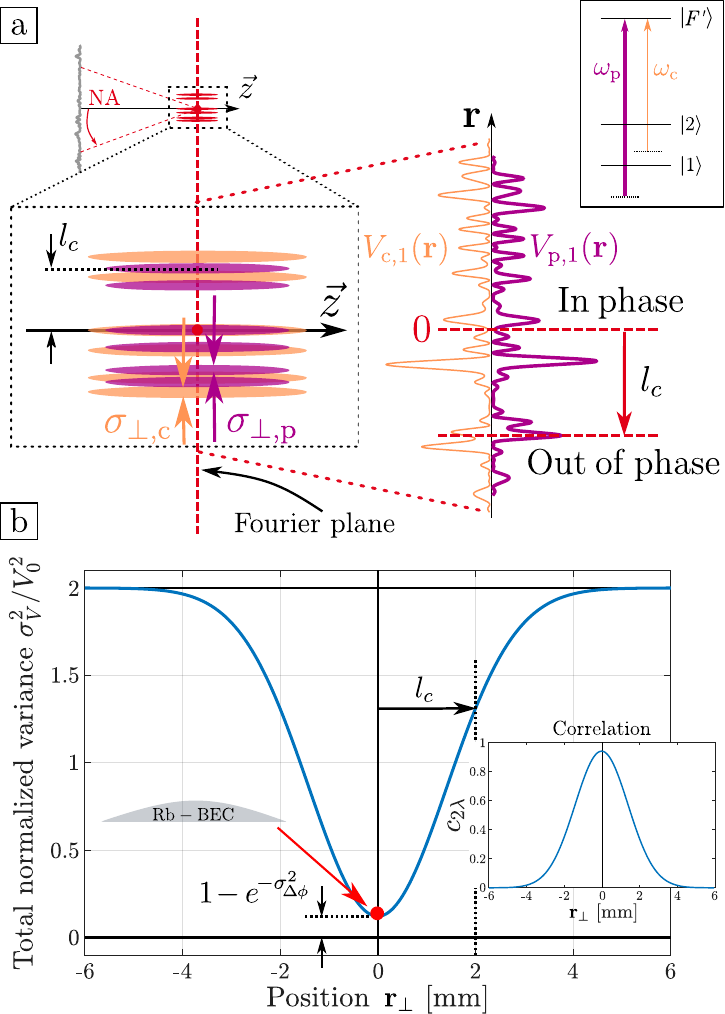}
\caption{a: Illustration of the correlation between two speckle potentials generated by the same diffuser with two slightly different wavelengths. Close to the Fourier plane, the two patterns are identical up to a spatial scaling factor, yielding an almost perfect overlap of the potential that decreases with the distance to the optical axis. The correlation length $l_c$ describes the typical distance for which the speckle patterns become out of phase. b: Plot of the variance of the total potential as a function of the position in the Fourier plane. Close to the optical axis, the two speckle patterns are similar and the only decorrelation term arises from the phase difference due to the propagation through the diffuser. Far away from the optical axis, the two speckle patterns do not superimpose and the variance of the total potential is the sum of the two individual variances. Plot obtained for a realistic inverse relative frequency difference $\mathcal{F}=4000$, a well-exaggerated diffuser's roughness $\sigma_\mathrm{e}^2=10^5 \lambda_\mathrm{p}^2$ and potentials of opposite average values $V_0$.}
\label{fig:sum_potentials}
\end{figure}

Relying on the same paraxial assumption as for the calculation of the spatial correlation function of a single speckle pattern, we find (see Appendix A) that the correlation function between the two speckles generated by the same diffuser, for the same spatial mode but for different frequencies, can be expressed as:
\begin{equation}
c_{2\lambda}(\mathbf{r}_\perp, z, \lambda_\mathrm{p}, \lambda_\mathrm{c}) = e^{-\sigma_{\Delta\phi}^2} \; c_{3D} \Big( \frac{\mathbf{r}_\perp}{\mathcal{F}}, \frac{ z}{\mathcal{F}} \Big) \,,
\label{eq:bichromatic_correlation}
\end{equation}
where $\sigma_{\Delta \phi}$ is a term discussed below and the parameter $\mathcal{F} = \omega_\mathrm{p}/\delta\omega = \lambda_\mathrm{p}/\delta\lambda\approx 4000$ characterizes the frequency difference of the two speckle fields. Expression (\ref{eq:bichromatic_correlation}) constitutes a major result of this letter.
It is the product of two terms that have a simple interpretation.

The first term $\exp{\lbrace-\sigma_{\Delta\phi}^2\rbrace}$ is associated to the fluctuations of the phase difference for the two different wavelengths $\lambda_p$ and $\lambda_c$ propagating inside the rough plate at the same point. More precisely, the phases at each point of the diffuser differ for the two lasers according to $\Delta\phi(\mathbf{r}_0) = \phi_\mathrm{p} - \phi_\mathrm{c} = 2\pi (n-1) (\lambda_\mathrm{p}^{-1} - \lambda_\mathrm{c}^{-1}) \delta e(\mathbf{r}_0)$, where $\delta e$ corresponds to the thickness fluctuations. When averaged over the diffuser, i.e. over $\mathbf{r}_\mathrm{0}$, and considering a Gaussian distribution for the phase as discussed in section~\ref{sc:section2_1}, one obtains an average phase factor $\overline{\exp{\lbrace i \Delta\phi\rbrace}} = \exp{\lbrace-\sigma_{\Delta\phi}^2/2\rbrace}$, with $\sigma_{\Delta\phi}^2$ the variance of $\Delta\phi$. This fluctuating phase difference term therefore results into a decorrelation factor constant in the speckle pattern. It is solely due to the roughness of the diffusing plate, and it can be rewritten as $\exp{\lbrace-\sigma_{\Delta\phi}^2\rbrace}\approx \exp{\lbrace-4\pi^2(n-1)^2\sigma_\mathrm{e}^2/ \lambda_\mathrm{p}^2 \mathcal{F}^2\rbrace}$~\footnote{The factor two compared to the first factor in Eq.~\eqref{eq:bichromatic_correlation} is due to the fact that we consider the correlation of the potential that is proportional to the light intensity and not to the field amplitude, see appendix~\ref{app:double_speckle}.}. Here $\sigma_\mathrm{e}$ is the thickness fluctuations r.m.s. value, i.e. the diffuser's roughness. It means that the rougher the diffuser and the bigger the frequency difference, the less the speckle patterns will be correlated, even at the center of the patterns. This term has been visually exaggerated in figure~\ref{fig:sum_potentials}.b and explains why the minimum of the variance does not reach perfectly 0.

In practice, the diffuser's roughness $\sigma_\mathrm{e}$ is fixed by manufacturing. It was measured at the optics workshop of Institut d'Optique, using a profilometer. Several measurements of the surface's profile have been performed along $\SI{1}{\milli\meter}$~long straight lines onto several areas of our diffuser, leading to a r.m.s. roughness of $\sigma_\mathrm{e}=\SI{1.3}{\micro\meter}$. For $\mathcal{F}\approx 4000$, the term $e^{-\sigma_{\Delta\phi}^2}$ (associated to the first factor of expression~\ref{eq:bichromatic_correlation}) is $\sim 1-10^{-6}$, 
indicating almost perfect correlation between the two speckles and thus a negligible residual disorder.

The second contribution in equation (\ref{eq:bichromatic_correlation}) describes the loss of correlation as the position in the speckle is shifted away from the optical axis. It is due to the different geometrical scaling factors of the two speckle patterns, proportional to the wavelengths, as illustrated in figure~\ref{fig:sum_potentials}. Remarkably, the resulting term is expressed with  the spatial correlation function (\ref{eq:c3D definition}) of a monochromatic speckle, with a magnifying factor $\mathcal{F}$. The bichromatic correlation function then has a width $l_c=\mathcal{F}\sigma_\perp$ which defines a ``bichromatic correlation length", corresponding to the distance from the center at which the two patterns are shifted by one speckle grain size. 

This correlation length must be compared to the maximum size of the atomic sample. Taking again $\mathcal{F}\approx 4000$ corresponding to a frequency difference  $\delta\omega/2\pi\approx \SI{100}{\giga\hertz}$ (see section~\ref{sc:bichromatic_principle}), the bichromatic correlation length is of the order of $l_c \sim \SI{2}{\milli\meter}$. It is much larger than the largest size of our atomic sample about $R_\mathrm{TF} \sim \SI{45}{\micro\meter}$, see section~\ref{sc:experiment}. We can then estimate this decorrelation factor to be at most
$1-c_{3D}(R_\mathrm{TF}/\mathcal{F}) \sim 5 \times10^{-4}$. Altogether, we find using Eq.~\eqref{eq:variance_bichromatic} that the fundamental potential decorrelation leads typically to a residual disorder in state $\lvert 1\rangle$ with a r.m.s. value of the order of
\begin{equation}
\sigma_{V_1} \sim 0.02 \times V_\mathrm{R} \, ,
\label{eq:residual_estimation}
\end{equation}
i.e. of the same order as in Volchkov \textit{et al.}~\cite{volchkov2018measurement}, see discussion in \ref{sc:section2_2_2}. For instance, we will see below that Table~\ref{tbl:mukhtar} predicts a residual disorder $\sigma_{V_1}/h = \SI{9.6}{\hertz}$ for the specific case of $V_\mathrm{R}/h=\SI{416}{\hertz}$, in agreement with the coarse estimation of expression~\eqref{eq:residual_estimation}.

\subsection{Reduction of the photon-scattering rate for the disorder-sensitive state}%%%%%%%%%%%%%%%%%%
\label{sc:lifetime}

Let us recall that a strong reduction of $\Gamma_2$, the photon scattering rate of the disorder-sensitive state $\lvert 2\rangle$, is the main goal of our present work. In order to study quantum transport phenomena such as Anderson localization, a lifetime on the order of one second time or more is needed~\cite{jendrzejewski2012three,semeghini2015measurement}. Note that  the condition is much less stringent in the initial state, being only on the order of tens of milliseconds,. This state $\lvert 1\rangle$ is indeed only used as a ``source" of atoms of very well defined energy during the rf transfer~\footnote{Once the rf transfer is switched off, the atoms in the state $\lvert 1\rangle$ can be removed to avoid any detrimental interactions with the atoms transferred in the state $\lvert 2\rangle$.}, yielding the simpler condition $\Gamma_1^{-1}\geq t_\mathrm{rf}$ (see section~\ref{sc:section2_2}).

In presence of the bichromatic potential, the total photon scattering rate is given by the sum of the individual rates, proportional to the imaginary parts of the atomic polarizability, in the linear regime:
\begin{eqnarray}
\Gamma_{F,m_F}&=&\frac{1}{\hbar\epsilon_0 c} \Big( \mathrm{Im}[\alpha_{F,m_F}(\delta_\mathrm{P})] I_\mathrm{p} \nonumber \\
&& \quad \quad \quad + \; \mathrm{Im}[\alpha_{F,m_F}(\delta_\mathrm{c})] I_\mathrm{c}\Big) \mathrm{ .}
\label{eq:lifetime_bichromatic}
\end{eqnarray}
Table \ref{tbl:mukhtar} shows the results of the numerical determination of the experimental parameters for a bichromatic speckle potential corresponding to $V_\mathrm{R}/h=\SI{416}{\hertz}$. This value is typical for the experimental study of the Anderson localization or the spectral functions (see Refs.~\cite{jendrzejewski2012three,volchkov2018measurement}). For consistency, this value will also corresponds to the experiments described in the next section.

\begin{table}[h]
\caption{Comparison of the state-dependent disordered potential parameter between the monochromatic configuration and the bichromatic one, the crucial quantity being the photon scattering lifetime in state $\lvert 2\rangle$ ($\Gamma_2^{-1}$, bolded line). The determination of these quantities has been performed for a total disorder amplitude of $V_\mathrm{R}/h=\SI{416}{\hertz}$.  The quantities $\delta_\mathrm{p}$ and $\delta_\mathrm{c}$ are defined in figure~\ref{fig:bichromatic} and $\delta$ is defined in figure~\ref{fig:monochromatic}. $P$, $P_\mathrm{p}$, and $P_\mathrm{c}$ correspond to the laser powers.}\label{tbl:mukhtar}%
\begin{center}
\begin{minipage}{0.46\textwidth}
 \begin{ruledtabular}
 \begin{tabular}{lll}
%\toprule
Quantity & Monochromatic case  & Bichromatic case  \\
\colrule
$\delta /2\pi$ & $\SI{81}{\mega\hertz}$ & \textemdash \\
$\delta_\mathrm{p}/2\pi$ & \textemdash & $\SI{95}{\giga\hertz}$  \\
$\delta_\mathrm{c}/2\pi$ & \textemdash & $\SI{-1.40}{\giga\hertz}$ \\
$\Gamma_1^{-1}$ & $\SI{26.6}{\second}$ & $\SI{73}{\milli\second}$ \\
$\mathbf{\Gamma_2^{-1}}$ & $\mathbf{5.3\:ms}$ & $\mathbf{1.66\:s}$ \\
$\sigma_{V_1}/h$ & $\SI{6.3}{\hertz}$ & $\SI{11.4}{\hertz}$ \\
$P$ & $\SI{0.49}{\micro\watt}$ & \textemdash\\
$P_\mathrm{p}$ & \textemdash & $\SI{430}{\micro\watt}$ \\
$P_\mathrm{c}$ &\textemdash & $\SI{4.6}{\micro\watt}$ \\
\end{tabular}
 \end{ruledtabular}
\end{minipage}
\end{center}
\end{table}

Here, the detuning of the principal laser is set to $\delta_\mathrm{p}/2\pi = \SI{95}{\giga\hertz}$. This value is chosen as a compromise to get a sufficiently large detuning while the two speckle patterns can be considered identical enough to permit the cancellation of the total potential onto the disorder insensitive state. Then, the detuning $\delta_\mathrm{c}$ of the compensation laser is determined so that the lifetime of state $\lvert 1\rangle$ is larger than the duration of rf transfer ( for instance  $t_\mathrm{rf}=\SI{20}{\milli\second}$ for the experiment shown in section~\ref{sc:experiment}). Here we have $\delta_\mathrm{c}/2\pi= -\SI{1.40}{\giga\hertz}$ and $\Gamma_1^{-1} = \SI{73}{\milli\second}$. With these parameters, one can in particular deduce the potential amplitude generated by both the laser speckle fields on the state $\lvert 1\rangle$ and
$\lvert 2\rangle$: $V_\mathrm{p,1}/h =- V_\mathrm{c,1}/h=\SI{366}{\hertz}$, $V_\mathrm{p,2}/h = \SI{348}{\hertz}$ and  $V_\mathrm{c,2}/h =\SI{68}{\hertz}$. 

The most notable result of Table \ref{tbl:mukhtar} is the improvement of the scattering lifetime $\Gamma_2^{-1}$ of the disorder-sensitive state $\lvert 2\rangle$ by more than two orders of magnitude, going from a few milliseconds for the monochromatic configuration of~\cite{volchkov2018measurement} up to more than one second in the bichromatic configuration. Additionally, the residual disordered potential $\sigma_{V_1}$ applied to the disorder-insensitive state is of the same order as in the monochromatic configuration, yielding the same state-selectivity. 

The analysis described here for the specific case of $V_\mathrm{R}/h=\SI{416}{\hertz}$ can be reproduced for a wide range of disorder amplitudes, both for a globally attractive ($V_\mathrm{R}<0$) and repulsive ($V_\mathrm{R}>0$) disorder in state $\lvert 2\rangle$, leading to the same conclusion. For instance, we checked that we obtain similar improvements for the range of disorder amplitudes $\lvert V_\mathrm{R}/h \rvert \in [ \SI{40}{\hertz},\, \SI{4}{\kilo\hertz} ] $ used in Volchkov \textit{et al.}~\cite{volchkov2018measurement}.

%%%%%%%%%%%%%%%%%%%%%%%%%%%%%%%%%%%%%%%%%%%%%%%%%%%%%%%
%%%%%%% SECTION 4: EXPERIMENTAL IMPLEMENTATION
%%%%%%%%%%%%%%%%%%%%%%%%%%%%%%%%%%%%%%%%%%%%%%%%%%%%%%%

\section{Experimental check of the bichromatic speckle configuration} 
\label{sc:experiment}

In this section we present an evaluation of the new bichromatic disordered potential scheme with ultracold atoms. More precisely, we focus on two main aspects. Firstly, we check the efficient subtraction of the two speckle potentials in the state $\lvert1\rangle$, as described in Fig.~\ref{fig:bichromatic}, so that it is merely insensitive to the disordered potential. This is done by studying the mechanical excitation of the atoms in state  $\lvert 1\rangle$ following a quench of the disorder potential. Secondly, we realize the rf transfer protocol as demonstrated in Volchkov~\textit{et al.}~\cite{volchkov2018measurement} (see Fig.~\ref{fig:principe}) and measure the spontaneous scattering lifetime of the atoms transferred in state $\lvert2\rangle$. As predicted in table~\ref{tbl:mukhtar}, we find a large improvement by two orders of magnitude compared to the monochromatic speckle case, the lifetime being now on the second time scale.

\subsection{Experimental set-up}
\label{sc:setup}

The starting point of the experiment is the creation of a BEC of $^{87}$Rb atoms in the hyperfine state $\lvert1\rangle \equiv \lvert F=1,m_{F}=-1\rangle$. The atoms are confined in a crossed optical dipole trap, formed by two orthogonal laser beams at a wavelength of 1070~nm, in presence of a magnetic levitation. As described in Ref.~\cite{volchkov2018sm}, the magnetic levitation is created by adding a magnetic gradient, whose force acts against gravity, to the magic bias field of 3.23 G (see section~\ref{sc:section2_2_1}). 

The magnetic levitation enables us to end up the optical evaporation process with a very decompressed trap configuration ($\omega_y \simeq \SI{5}{\hertz}, \,        \omega_z / 2 \pi  \simeq \SI{25}{\hertz}$ and $\omega_x / 2 \pi  \simeq \SI{30}{\hertz}$). At this stage, we obtain a BEC with around 2$\times$10$^{5}$ atoms, corresponding to a chemical potential around $\mu/h \simeq \SI{250}{\hertz}$ and a Thomas-Fermi radii around $R_\mathrm{TF} \sim$ 45, 10 and 8 $\SI{}{\micro\meter}$ along each direction. In the Thomas-Fermi regime, the meanfield interatomic interactions compensate perfectly the trapping potential~\cite{volchkov2018sm}, resulting in an overall flat potential for atoms in state $\lvert 1\rangle$ and in state $\lvert 2\rangle$ in the absence of  the disordered potential, as sketched in Figs~\ref{fig:principe},\ref{fig:monochromatic} and \ref{fig:bichromatic}.

The laser speckle disorder geometry and parameters have been extensively detailed in sections \ref{sc:section2_1} and \ref{sc:section3}. Let us note, however, that the precise disorder amplitude calibration is a well-known issue for experiments. As shown in~\cite{volchkov2018sm}, an efficient method relies on direct comparison between directly measured spectral functions and numerical calculations. The same method is thus applied here using the spectral function obtained for our reference disorder amplitude value $V_\mathrm{R}/h=\SI{416}{\hertz}$ (see section~\ref{sc:Improvement of scattering lifetime}).

\subsection{Probing the insensitivity of state $\lvert 1\rangle$ to disorder using a quench}
\label{sc:quench}

In order to quantify the effect of the residual disorder in state $\lvert 1\rangle$, we introduce a quench protocol. To do so, we switch on abruptly the total disordered potential (within $\SI{100}{\micro\second}$) and keep it on for $\SI{4}{\milli\second}$. Then the disorder is switched off and we measure the momentum distribution of the atoms in $\lvert 1\rangle$ using a standard time of flight technique of duration $t_\mathrm{ToF}=\SI{200}{\milli\second}$. 

\begin{figure}
\centering
\includegraphics[width=0.45\textwidth]{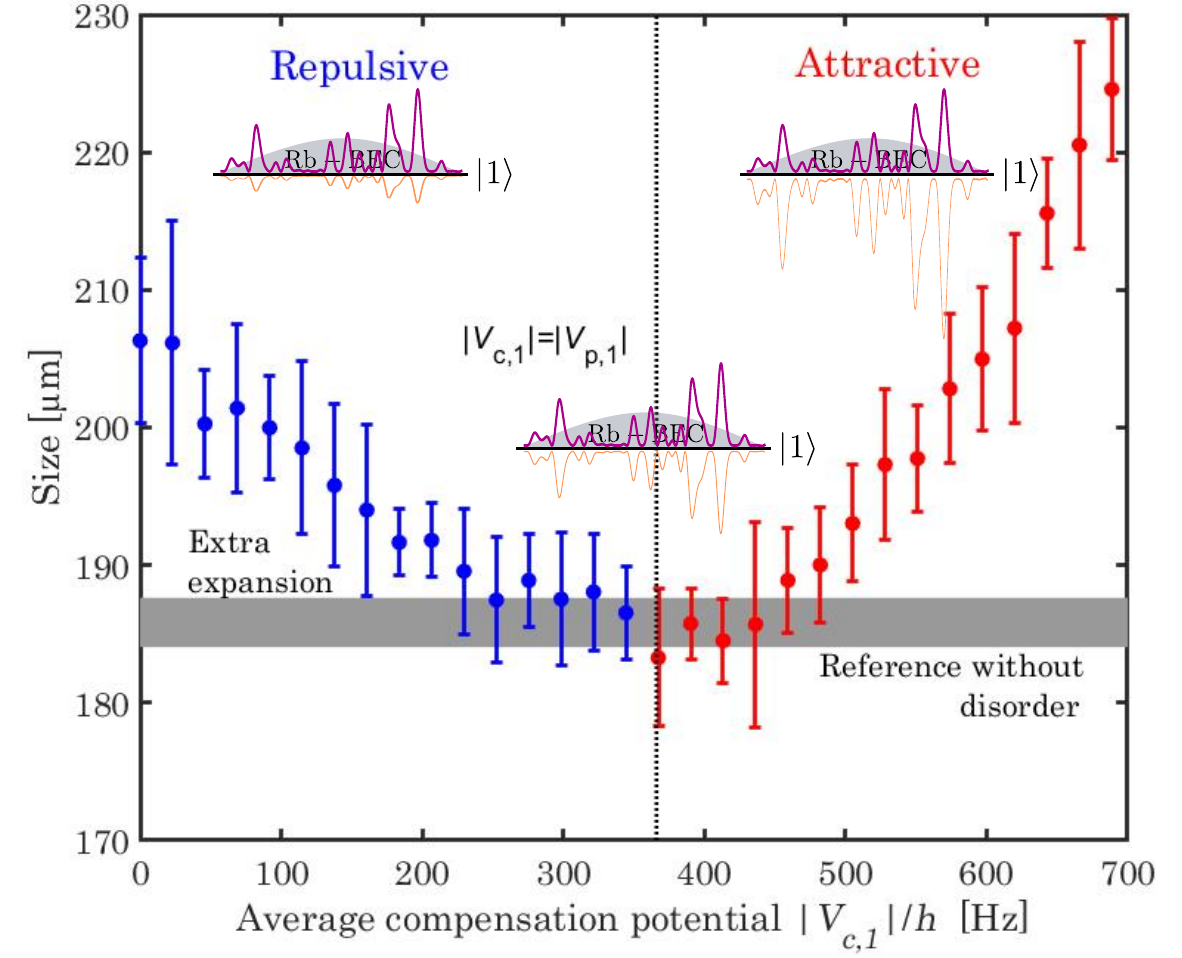}
\caption{Evolution of the momentum distribution of the atoms in state $\lvert 1\rangle$ following a sudden switch on of the bichromatic speckle potential (quench protocol). The amplitude of the principal potential amplitude is fixed to $V_\mathrm{p,1}/h=\SI{366}{\hertz}$, while the amplitude of compensating potential is scanned from $\lvert V_\mathrm{c,1} \rvert/h$=0 to $ \SI{700}{\hertz}$. The dots correspond to the atomic cloud r.m.s. size measured by fluorescence imaging after a time of flight of 200~ms. The error bars correspond to the rms uncertainties estimated over five experimental realizations. For clarity the dots are colored in blue when the total disorder amplitude $V_{1}=V_\mathrm{p,1}+V_\mathrm{c,1}$ is repulsive and in red when it is attractive -- see illustrations. The horizontal line corresponds to the reference case where no disorder is applied to the atoms. The vertical dotted line indicates the theoretical condition for an optimal cancellation of the total disorder potential in state $\lvert 1\rangle$, that is $\lvert V_\mathrm{c,1}\rvert$ = $V_\mathrm{p,1}$. }
\label{fig:compensation}
\end{figure}

The evolution of the atomic momentum distribution is shown on figure~\ref{fig:compensation}, where we scan the amplitude of the compensation potential from $\lvert V_\mathrm{c,1}\rvert/h =0$ to $\SI{700}{\hertz}$ ($V_\mathrm{c,1}$ being attractive), while the principal potential is kept constant $V_\mathrm{p,1}/h=\SI{366}{\hertz}$ ($V_\mathrm{p,1}$ being repulsive). These parameters are chosen to explore the typical configuration predicted in table~\ref{tbl:mukhtar}, where the state $\lvert 2\rangle$ experiences the total potential $V_\mathrm{R}/h=\SI{416}{\hertz}$ while the disorder is ideally suppressed in state $\lvert 1\rangle$ for $ V_\mathrm{c,1}= - V_\mathrm{p,1}$ (see section~\ref{sc:cancellation_potential}).

For very low compensation amplitude, $\lvert V_\mathrm{c,1}\rvert  \ll V_\mathrm{p,1}$, the total potential in state $\lvert 1\rangle$ is essentially given by the ``principal" disorder potential. This repulsive potential excites the BEC when the disorder is switched on, a part of the potential energy being transfer to the kinetic energy. This extra kinetic energy results in a broadening of the atomic momentum distribution (see dashed horizontal line corresponding to the absence of disorder as a reference). 

As the compensating potential is increased, the total disordered potential in $\lvert 1\rangle$ decreases and the momentum distribution spread decreases accordingly, with a minimum at the best compensation. When $\lvert V_\mathrm{c,1} \rvert > V_\mathrm{p,1} $, the total potential $V_{1}$ turns to attractive, and the disorder strength increases again. The momentum distribution broadens then in this regime. The minimum is reached as expected around $\lvert V_\mathrm{c,1} \rvert$ = $ V_\mathrm{p,1} $ (vertical thin dotted line), and, most importantly we observe a momentum spread exactly equal  to the one in the absence of disorder. This observation is a strong evidence of the efficient compensation  of the two disordered potentials, the residual potential yielding no observable excitation of the atomic cloud in $\lvert 1\rangle$.

\subsection{Improved lifetime in the disorder sensitive state $\lvert 2\rangle$}
\label{sc:Improvement of scattering lifetime}

The expected suppression of the disordered potential in state $\lvert 1 \rangle$ being verified, we perform the rf transfer  towards the disorder sensitive state  $\lvert 2 \rangle$ at the  energy defined by $\delta_\mathrm{rf}= \omega_\mathrm{rf}- \Delta_\mathrm{hf}$ (see Fig.~\ref{fig:principe} and discussion in section~\ref{sc:section2_2_1}). Here we continue to investigate the configuration of Table~\ref{tbl:mukhtar} and we set the bichromatic disorder parameters at the minimum point of figure~\ref{fig:compensation}, that is for $\lvert V_\mathrm{c,1}\rvert /h$ = $\ V_\mathrm{p,1} /h = \SI{366}{\hertz}$. The rf power is chosen low enough  to operate in the weak coupling regime where the transfer rate $\Gamma(\delta_\mathrm{rf})$ is well predicted by the Fermi Golden rule~\cite{volchkov2018sm}. Moreover, the rf field is applied in the regime of $\Gamma t_\mathrm{rf} \ll 1$ (with  $t_\mathrm{rf}=\SI{20}{\milli\second}$) so that only a small fraction of the atoms are transferred to state $\lvert 2\rangle$ (not more than $25\%$ at most). As discussed in section~\ref{sc:lifetime}, note that the rf duration is also chosen to be shorter than the lifetime in state $\lvert 1 \rangle$. In these conditions, the energy resolution is time Fourier-limited to $\Delta E/h=1/t_\mathrm{rf}= \SI{50}{\hertz}$. 

\begin{figure}
\centering
\includegraphics[width=0.4\textwidth]{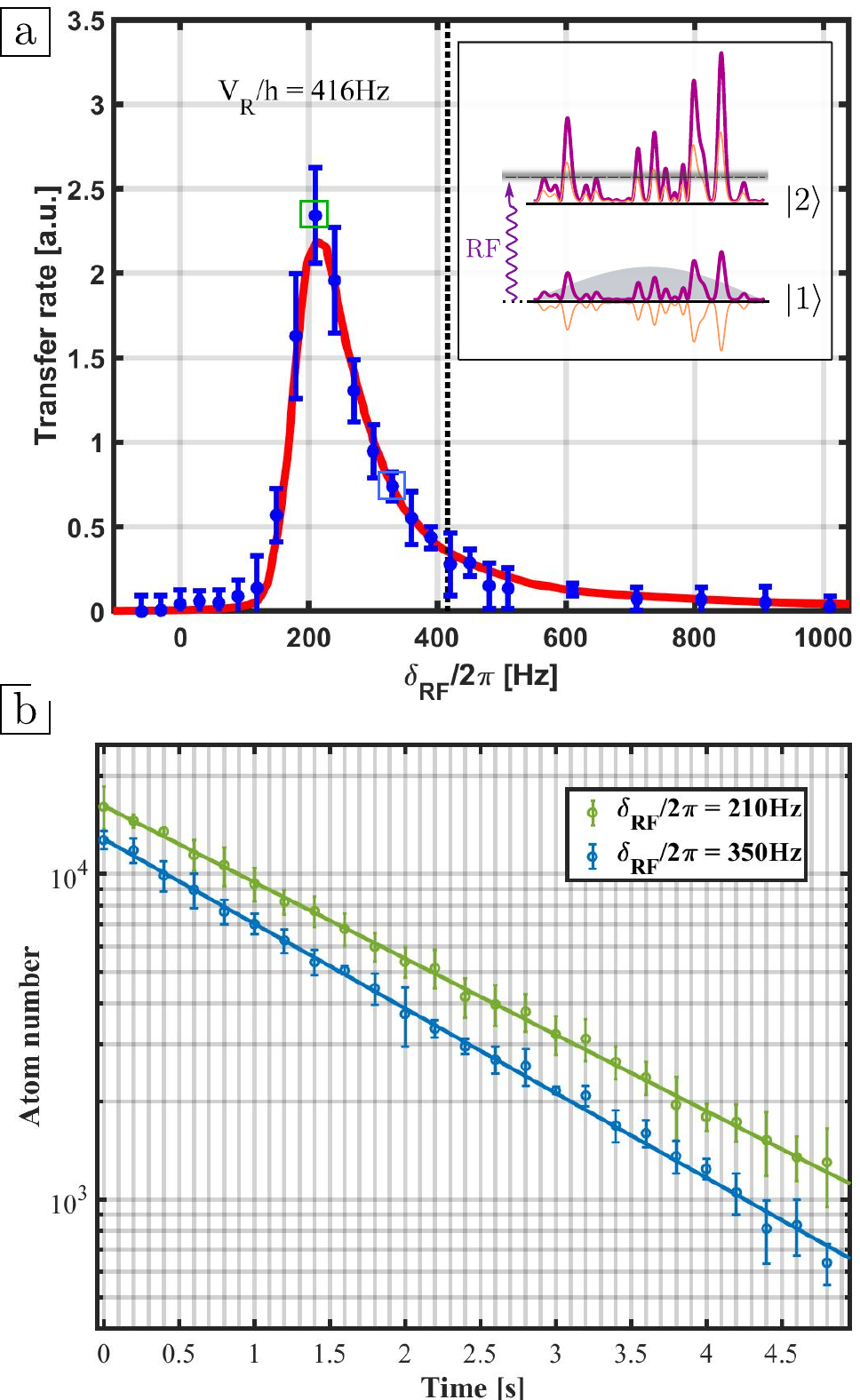}
\caption{Implementation of the rf transfer scheme and lifetime measurement for $ V_\mathrm{R}/h=\SI{416}{Hz}$. a) Normalized transfer rate $\Gamma(\delta_\mathrm{rf})$ from disorder insensitive state $\lvert1\rangle$ to the disorder sensitive state  $\lvert2\rangle$. The compensating and principal disorder amplitudes are set to $\lvert V_\mathrm{c,1}\rvert /h$ = $\ V_\mathrm{p,1} /h = \SI{366}{\hertz}$, that is the optimal cancellation condition for the disorder in state $\lvert1\rangle$ (see Fig.~\ref{fig:compensation}). The rf power is fixed and the rf field is applied during $t_\mathrm{rf}=\SI{20}{\milli\second}$. The blue dots are the measured points and the red curve is the numerical calculation in \cite{volchkov2018measurement}, taking into account the resolution $\Delta E/h=1/t_\mathrm{rf}=\SI{50}{\hertz}$. The squares corresponds to the detunings $\delta_\mathrm{rf}$ chosen to measure the lifetime.  b) Atom number decays on the state $\lvert 2\rangle $ after the transfer for the detunings $\delta_\mathrm{rf}/2\pi=210$ and $\SI{350}{\hertz}$. The fits with exponential decays yield the lifetime of $\SI{1.85 \pm 0.05}{\second}$ and $\SI{1.67 \pm 0.06}{\second}$ respectively, in good agreement with Table~\ref{tbl:mukhtar}. For both figures, the error bars correspond to the rms uncertainties estimated over five experimental realizations.  }
\label{fig:transfer and lifetime}
\end{figure}

Figure~\ref{fig:transfer and lifetime}a shows the evolution of the transferred atom number as a function of the rf detuning $\delta_\mathrm{rf}$, for a fixed rf power. As explained in~\cite{volchkov2018measurement}, the curve $\Gamma(\delta_\mathrm{rf})$ constitutes a direct measurement of the spectral function for the disordered potential in state  $\lvert 2 \rangle$. A comparison is shown with a numerical simulation for a repulsive disorder of amplitude $V_\mathrm{R}/h=\SI{416}{\hertz}$. As said above this excellent agreement is used to calibrate precisely the disorder amplitude in the experiments (with a 5$\%$ uncertainty).

Last we study the scattering lifetime of the atoms transferred in the disorder sensitive state $\lvert 2\rangle$ for the same configuration corresponding to $V_\mathrm{R}/h = \SI{416}{\hertz}$. In order to maximise the signal to noise ratio we choose two rf detuning, $\delta_\mathrm{rf}/2\pi=\SI{210}{\hertz}$ and $\SI{350}{\hertz}$, close to the maximum of the transfer curve $\Gamma(\delta_\mathrm{rf})$. The decay of the atoms number (once the rf transfer field is switched off) are shown on figure~\ref{fig:transfer and lifetime}b). The fits by exponential curves yield very similar lifetime of respectively $\tau_{210}= \SI{1.85 \pm 0.05}{\second}$ and $\tau_{350}= \SI{1.67 \pm 0.06}{\second}$. These values are in very good agreement with the prediction in table~\ref{tbl:mukhtar}, thus validating our analysis. Most importantly these values are larger than one second, which is crucial for our ongoing investigation of the Anderson transition.

%%%%%%%%%%%%%%%%%%%%%%%%%%%%%%%%%%%%%%%%%%%%%%%%%%%%%%%
%%%%%%% SECTION 5: CONCLUSION
%%%%%%%%%%%%%%%%%%%%%%%%%%%%%%%%%%%%%%%%%%%%%%%%%%%%%%%
\section{Summary and outlook}

In this paper, we have studied both theoretically and experimentally a new scheme to produce a 3D state dependent disordered potential with a low spontaneous photon scattering rate in the disordered sensitive state. It is realized using a bichromatic speckle disorder whose properties have been investigated in details, in particular to quantify the conditions under which the two disordered potentials, created from two laser speckle field at two slightly different wavelengths, can cancel efficiently together for the insensitive state. 
Using this state dependent disorder potential and a rf transfer from the insensitive state to the disorder sensitive state, one can load ultracold atoms at a precise energy level in the disordered potential. The transferred atoms have a typical photon scattering lifetime on the second time scale, an improvement of two orders of magnitude compared to the monochromatic speckle scheme~\cite{volchkov2018measurement}. We therefore expect that scheme to allow us to effect precise measurements of the mobility edge and the critical exponents the 3D Anderson transition, whose precise determination remains an utmost experimental challenge~\cite{pasek2017anderson}. 

It is worth noting that this scheme could be extended to other kind of optical potentials used to study disordered systems, such as those created by spatial light modulators as well as quasi-periodic lattices~\footnote{While the general arguments exposed in section~\ref{sc:cancellation_potential} on the fundamental decorrelation of the potentials produced by two different wavelength would be essentially similar, each configuration would require specific calculations. A general analysis is, however, beyond the scope of this paper.}, see e.g. \cite{roati2008,choi2016}. It opens also many prospects to address fundamental questions related to the Anderson transition, such as the observation of multifractalilty~\cite{werner2018multifractal}, comparison with new theoretical approaches such as the one based on the ``hidden landscape"~\cite{filoche2012,pelletier2022spectral}, or general Anderson transition with different universality class  \cite{wang2021anderson}, dimensions~\cite{orso2017transition}, or even in the presence of non-hermitian disorder~\cite{Luo2021anderson}.

\begin{acknowledgments}
The authors thanks Vincent Denechaud and Adrien Signoles for early discussions and work. This work was supported by grant No. 601937 from the Simons Foundation and by an ``Investissements d'Avenir" grant from LabEx PALM (ANR-10-LABX-0039-PALM). A.A. acknowledges support through the Augustin Fresnel chair of the Institut d'Optique Graduate School, sponsored by Institut d'Optique and supported by Nokia Bell labs. He also acknowledges support from the iXcore-iXlife-IXblue foundation for research.
\end{acknowledgments}
%
%\section*{Declarations}
%
%\begin{itemize}
%\item  All authors contributed equally to the paper.
%
%
%\item The authors declare that they have no conflict of interest.
%
%
%\item  The datasets generated during and/or analyzed during the current study are available from the corresponding author on reasonable request.
%
%%\item Consent to participate
%%\item Consent for publication
%%\item Code availability 
%\end{itemize}

%%%%%%%%%%%%%%%%%%%%%%%%%%%%%%%%%%%%%%%%%%%%%%%%%%%%%%%
%%%%%%% SECTION A: APPENDIX
\appendix

\section{Calculation of the bichromatic correlation function}
\label{app:double_speckle}

\subsection{Bichromatic correlation function of the diffuser}  
\label{A1}

We extend the standard calculation of the diffuser correlation function to the bichromatic case. The laser's local phase fluctuations just after the diffuser is expressed as:
\begin{equation}
\phi(\mathbf{r}_\mathrm{0})=2\pi (n-1) \frac{\delta e(\mathbf{r}_\mathrm{0})}{\lambda}
\end{equation}
where the points $\mathbf{r}_\mathrm{0}=\lbrace x_0, y_0, z=0\rbrace$ correspond to the diffuser's plane. $\delta e(\mathbf{r}_\mathrm{0})$ is the local fluctuation of the thickness of the diffuser, $\lambda$ the laser wavelength and $n$ stands for the glass index. Note that within this definition, we don't consider the averaged phase transmission, that we set to $\overline{\phi}=0$ for simplicity. Moreover, we assume the diffuser to have a gaussian-distributed thickness with standard deviation $\sigma_\mathrm{e}$, resulting in a phase standard deviation $\sigma_{\phi} = 2\pi (n-1) \sigma_\mathrm{e} /\lambda$. The diffuser transmission is $t_\mathrm{diff}(\mathbf{r}_\mathrm{0},\lambda)=e^{i \phi(\mathbf{r}_\mathrm{0})}$, and its ensemble average writes :  
\begin{equation}
\overline{t_\mathrm{diff}}= \overline{e^{i \phi}} = \int{ \mathrm{d}\phi \: e^{i \phi} \: \mathcal{P}(\phi)} = e^{-\sigma_{\phi}^2/2} \label{eq:property_gaussian_variable}
\end{equation}
with $\mathcal{P}(\phi) = 1/\sqrt{2 \pi}\sigma_{\phi} \times  \exp{(-\phi^2 /2 \sigma_{\phi}^2)}$.

The bichromatic correlation function of the diffuser is defined as:
\begin{equation}
C_\mathrm{diff}(\mathbf{r}_\mathrm{0},\mathbf{r}_\mathrm{0}',\lambda_\mathrm{p},\lambda_\mathrm{c})=\overline{t_\mathrm{diff}(\mathbf{r}_\mathrm{0},\lambda_\mathrm{p}) t_\mathrm{diff}^*(\mathbf{r}_\mathrm{0}',\lambda_\mathrm{c})} \: .
\end{equation}
Note that the correlation function of the diffuser with a monochromatic illumination is obtained by merely setting $\lambda_\mathrm{c}=\lambda_\mathrm{p}$. Since the diffuser's thickness is a gaussian random variable, the phase difference $\phi_\mathrm{p}(\mathbf{r}_\mathrm{0})-\phi_\mathrm{c}(\mathbf{r}_\mathrm{0}')$ is a gaussian variable as well and we can calculate as in equation (\ref{eq:property_gaussian_variable}) to obtain:
\begin{align}
&C_\mathrm{diff}(\mathbf{r}_\mathrm{0},\mathbf{r}_\mathrm{0}',\lambda_\mathrm{p},\lambda_\mathrm{c}) = \overline{e^{i(\phi_\mathrm{p}(\mathbf{r}_\mathrm{0}) - \phi_\mathrm{c}(\mathbf{r}_\mathrm{0}'))}} \\
&= \exp{\left[-2\pi^2 (n-1)^2 \sigma_\mathrm{e}^2 \left( \frac{1}{\lambda_\mathrm{p}^2} + \frac{1}{\lambda_\mathrm{c}^2}\right) \right]} \nonumber \\
&\times \exp{\left[ 4\pi^2 (n-1)^2 \frac{2}{\lambda_\mathrm{p}\lambda_\mathrm{c}} \overline{\delta e(\mathbf{r}_\mathrm{0})\delta e(\mathbf{r}_\mathrm{0}')}\right]} \: .
\end{align}
To characterize the granularity correlation function  $\overline{\delta e(\mathbf{r}_\mathrm{0})\delta e(\mathbf{r}_\mathrm{0}')}$ of the diffuser, we introduce
$r_\mathrm{e}$, which is the width of the thickness correlation function and describes the typical size of the transverse granularity of the diffuser's surface. By assuming a wide phase distribution $\sigma_{\phi} \gg 2\pi$ (equivalently $\sigma_\mathrm{e} \gg \lambda$), the wave oscillates many times in a single granularity and the diffuser's thickness correlation function can be approximated by a bell-shaped curve:
\begin{equation}
\frac{\overline{\delta e(\mathbf{r}_\mathrm{0})\delta e(\mathbf{r}_\mathrm{0}')}}{\sigma_\mathrm{e}^2}\approx 1- \frac{(\mathbf{r}_\mathrm{0} - \mathbf{r}_\mathrm{0}')^2}{2 r_\mathrm{e}^2}
\end{equation}
when $ \left\lvert \mathbf{r}_\mathrm{0}-\mathbf{r}_\mathrm{0}' \right\rvert \ll r_\mathrm{e}$.
Finally, the bichromatic diffuser correlation function  reads as:
\begin{align}
C_\mathrm{diff}(\mathbf{r}_\mathrm{0},\mathbf{r}_\mathrm{0}',\lambda_\mathrm{p},\lambda_\mathrm{c}) 
&= \exp{\left( -\frac{\sigma_{\Delta\phi}^2}{2} \right)} \nonumber \\
 &\times \exp{\left( -\frac{(\mathbf{r}_\mathrm{0}-\mathbf{r}_\mathrm{0}')^2}{2 r_\mathrm{diff,p} r_\mathrm{diff,c}}\right)}
\end{align}
where $\sigma_{\Delta\phi}^2=(\sigma_{\phi_\mathrm{p}} - \sigma_{\phi_\mathrm{c}})^2$ is the variance of the local phase difference $\Delta\phi(\mathbf{r}_\mathrm{0}) = \phi_\mathrm{p}(\mathbf{r}_\mathrm{0}) - \phi_\mathrm{c}(\mathbf{r}_\mathrm{0})$ and where we introduced the diffuser's phase correlation length $r_\mathrm{diff,p,c} = r_\mathrm{e} / \sigma_{\phi, p,c}$ (respectively for the principal and compensating lasers).

\subsection{Correlation function of the amplitude} \label{A2}
We consider here the general two-point correlation function of bichromatic speckle field, at any positions $\mathbf{r}$ and $\mathbf{r}' $ and for different wavelengths $\lambda_\mathrm{p}$ and $\lambda_\mathrm{c}$. Within the paraxial approximation, it is expressed as
\begin{align}
&\Gamma_\mathcal{A}(\mathbf{r}, \mathbf{r}', \lambda_\mathrm{p}, \lambda_\mathrm{c}) \nonumber
= \overline{\mathcal{A}(\mathbf{r},\lambda_\mathrm{p})\mathcal{A}^*(\mathbf{r}',\lambda_\mathrm{c})}\\
&= \frac{e^{ik_\mathrm{p} \left( \overline{z}+\frac{x^2+y^2}{2\overline{z}} \right)}e^{-i k_\mathrm{c} \left( \overline{z}'+ \frac{x'^2+y'^2}{2\overline{z}'} \right)}}{\lambda_\mathrm{p}\lambda_\mathrm{c} \overline{z}\overline{z}'}\nonumber \\
\nonumber & \times \displaystyle\int{\mathrm{d}\mathbf{r}_\mathrm{0} \mathrm{d}\mathbf{r}_\mathrm{0}' \: \overline{t_\mathrm{diff}(\mathbf{r}_\mathrm{0}, \lambda_\mathrm{p})t_\mathrm{diff}^*(\mathbf{r}_\mathrm{0}', \lambda_\mathrm{c})}} \\
&\nonumber \times \mathcal{A}(\mathbf{r}_\mathrm{0},\lambda_\mathrm{p}) \mathcal{A}^*(\mathbf{r}_\mathrm{0}',\lambda_\mathrm{c}) \\
&\times e^{ik_\mathrm{p}\frac{ \mathbf{r}_\mathrm{0}^2}{2d_\mathrm{eff}}} e^{-ik_\mathrm{c}\frac{ \mathbf{r}_\mathrm{0}'^2}{2d_\mathrm{eff}'}} e^{-ik_\mathrm{p} \frac{\mathbf{r} . \mathbf{r}_\mathrm{0}}{\overline{z}}} e^{ik_\mathrm{c} \frac{\mathbf{r}' . \mathbf{r}_\mathrm{0}'}{\overline{z}'}} 
\end{align}
where $\mathcal{A}( \mathbf{r}_0)$ is the amplitude just before the diffuser, and the ensemble average $\overline{\:\cdots\:}$ only acts on the transmissions $t_\mathrm{diff}$. The effective distances $d_\mathrm{eff}$ are defined with respect to the focal plane by $1/d_\mathrm{eff} = 1/\overline{z}-1/f$. Last $\overline{z}=z+f$ is the longitudinal distance compared to the diffuser ($z$ being the distance to the Fourier plane, see Fig.~\ref{fig:geometry_speckle}).

Using the change of variables $\lbrace \mathbf{r}_\mathrm{0},\mathbf{r}_\mathrm{0}' \rbrace\rightarrow \lbrace \mathbf{r}_\mathrm{c,0}=(\mathbf{r}_\mathrm{0}+\mathbf{r}_\mathrm{0}')/2 ,\: \Delta\mathbf{r}_0=\mathbf{r}_\mathrm{0}'-\mathbf{r}_\mathrm{0} \rbrace$, we obtain:
\begin{align}
&\Gamma_\mathcal{A}(\mathbf{r}, \mathbf{r}', \lambda_\mathrm{p}, \lambda_\mathrm{p}) \propto \frac{e^{ik_\mathrm{p}\left( \overline{z} + \frac{x^2+y^2}{2\overline{z}} \right)} e^{-ik_\mathrm{c} \left( \overline{z}' + \frac{x'^2+y'^2}{2\overline{z}'}\right)}}{\lambda_\mathrm{p} \lambda_\mathrm{c} \overline{z} \overline{z}'} \nonumber \\
\nonumber & \times \displaystyle\int{\mathrm{d}\mathbf{r}_\mathrm{c,0} \mathrm{d}\Delta\mathbf{r}_0 \: C_\mathrm{diff}(\Delta\mathbf{r}_0,\lambda_\mathrm{p},\lambda_\mathrm{c})} \\
&\nonumber \times\mathcal{A}(\mathbf{r}_\mathrm{c,0}-\Delta\mathbf{r}_0/2,\lambda_\mathrm{p}) \mathcal{A}^*(\mathbf{r}_\mathrm{c,0}+\Delta\mathbf{r}_0/2,\lambda_\mathrm{c})  \\
&\nonumber \times   e^{i\mathbf{r}_\mathrm{c,0}.(\frac{k_\mathrm{c} \mathbf{r}'}{\overline{z}'} - \frac{k_\mathrm{p} \mathbf{r}}{\overline{z}})} e^{i\frac{\Delta\mathbf{r}_0}{2} . (\frac{k_\mathrm{p} \mathbf{r}}{\overline{z}} + \frac{k_\mathrm{c} \mathbf{r}'}{\overline{z}'})} e^{i\frac{\mathbf{r}_\mathrm{c,0}^2}{2} (\frac{k_\mathrm{p}}{d_\mathrm{eff}} - \frac{k_\mathrm{c}}{d_\mathrm{eff}'})}\\
& \times  e^{-i \frac{\Delta\mathbf{r}_0 . \mathbf{r}_\mathrm{c,0}}{2} ( \frac{k_\mathrm{p}}{d_\mathrm{eff}} + \frac{k_\mathrm{c}}{d_\mathrm{eff}'})} e^{i \frac{\Delta\mathbf{r}_0^2}{8}(\frac{k_\mathrm{p}}{d_\mathrm{eff}} - \frac{k_\mathrm{c}}{d_\mathrm{eff}'})} \: .
\end{align}

Let us now suppose that the typical size of the diffuser grains is small compared to the size of the incoming illumination $I(\mathbf{r}_0)$, meaning that on the size of $C_\mathrm{diff}$, the incoming illumination is constant. Let us also suppose that the two incoming beams are in the same spatial mode: $\mathcal{A}(\mathbf{r}_\mathrm{0}-\Delta\mathbf{r}_0/2,\lambda_\mathrm{p}) \mathcal{A}^*(\mathbf{r}_\mathrm{0}+\Delta\mathbf{r}_0/2,\lambda_\mathrm{c}) \approx \mathcal{A}(\mathbf{r}_\mathrm{0},\lambda_\mathrm{p}) \mathcal{A}^*(\mathbf{r}_\mathrm{0},\lambda_\mathrm{c}) = I(\mathbf{r}_\mathrm{0})$ (for simplicity we discard from now the c subscript from $\mathbf{r}_\mathrm{c,0}$). We find:
\begin{align}
&\Gamma_\mathcal{A}(\mathbf{r}, \mathbf{r}', \lambda_\mathrm{p}, \lambda_\mathrm{c}) = \frac{e^{ik_\mathrm{p}\left( \overline{z} + \frac{x^2+y^2}{2\overline{z}} \right)} e^{-ik_\mathrm{c} \left( \overline{z}' + \frac{x'^2+y'^2}{2\overline{z}'}\right)}}{\lambda_\mathrm{p} \lambda_\mathrm{c} \overline{z} \overline{z}'} \nonumber\\
&\nonumber \times \displaystyle\int{\mathrm{d}\mathbf{r}_\mathrm{0} \: I(\mathbf{r}_\mathrm{0}) e^{i\frac{\mathbf{r}_\mathrm{0}^2}{2}(\frac{k_\mathrm{p}}{d_\mathrm{eff}} - \frac{k_\mathrm{c}}{d_\mathrm{eff}'})} e^{i \mathbf{r}_\mathrm{0}. (\frac{k_\mathrm{c} \mathbf{r}'}{\overline{z}'}-\frac{k_\mathrm{p} \mathbf{r}}{\overline{z}})}} \\
&\nonumber \times \displaystyle\int{\mathrm{d}\Delta\mathbf{r}_0 \: C_\mathrm{diff}(\Delta\mathbf{r}_0,\lambda_\mathrm{p},\lambda_\mathrm{c}) \: e^{-i\frac{\Delta\mathbf{r}_0 . \mathbf{r}_\mathrm{0}}{2}(\frac{k_\mathrm{p}}{d_\mathrm{eff}} + \frac{k_\mathrm{c}}{d_\mathrm{eff}'})}} \\
&\times e^{i\frac{\Delta\mathbf{r}_0}{2}. (\frac{k_\mathrm{p} \mathbf{r}}{\overline{z}} +\frac{k_\mathrm{c} \mathbf{r}'}{\overline{z}'})} \: .
\end{align}

\subsection{3D monochromatic correlation close to the Fourier plane}  \label{A3}

In this part, we give the explicit calculation of the \textit{spatial} autocorrelation function of a monochromatic laser speckle field, i.e., at two different points $\mathbf{r},\mathbf{r}'$ but at the same wavelength $\lambda_\mathrm{p}=\lambda_\mathrm{c}=\lambda$. Such derivation is very well-known~\cite{goodman2007speckle} but it will be used for the derivation of the bichromatic correlation function in a simple form, which is an important result of the paper. The two point autocorrelation function is defined as 
\begin{equation}
c_{\mathrm{3D}}(\Delta\mathbf{r}_{\perp},\Delta z)=\frac{\overline{\delta I(\mathbf{r}_{\perp} + \Delta\mathbf{r}_{\perp}, z+\Delta z) \delta I(\mathbf{r}_{\perp}, z)}}{\overline{\delta I^2}}
\end{equation}
with $\mathbf{r}_\perp = \lbrace x,y\rbrace$. Close to the Fourier plane and to the optical axis, the numerator reads as $\overline{\delta I(\Delta\mathbf{r}_{\perp},\Delta z) \delta I(\mathbf{0}, 0)}$ and is computed using Wick's theorem:
\begin{align}
\label{eq:cal_mono_corr}
&\overline{\delta I (\Delta\mathbf{r}_{\perp}, \Delta z) \delta I(\mathbf{0},0)} = \lvert\Gamma_\mathcal{A}(\Delta\mathbf{r}_\perp, \Delta z)\rvert^2 \nonumber\\
&= \left\lvert \overline{\mathcal{A}(\Delta\mathbf{r}_{\perp},\Delta z) \mathcal{A}^*(\mathbf{0},0)} \right\rvert^2 \nonumber \\
&= \left\lvert \frac{1}{\lambda^2 f^2} \displaystyle\int{\mathrm{d}\mathbf{r}_\mathrm{0} \: I(\mathbf{r}_\mathrm{0}) \: e^{-i\frac{\mathbf{r}_\mathrm{0}^2 k \Delta z}{2f^2}} \: e^{-i\frac{k \mathbf{r}_\mathrm{0}. \Delta\mathbf{r}_{\perp}}{f}}}\right.\nonumber \\
&\times \left.\displaystyle\int{\mathrm{d}\Delta\mathbf{r}_0 \: C_\mathrm{diff}(\Delta\mathbf{r}_0) \: e^{i\frac{\Delta\mathbf{r}_0 . \mathbf{r}_\mathrm{0} k \Delta z}{2 f^2}} \: e^{i\frac{k\Delta\mathbf{r}_0 . \Delta\mathbf{r}_{\perp}}{2f}}} \right\rvert^2   \: .
\end{align}
By comparing the different phase terms, we can neglect the exponentials of the second integral. After normalization by $\lvert\overline{I(\mathbf{0},0)}\rvert^2$, and identifying a Fourier transform, we obtain the usual result~\cite{goodman2007speckle}: 
\begin{equation}
c_{\mathrm{3D}}(\Delta \mathbf{r}_{\perp} , \Delta z) = \frac{{\left\lvert \mathrm{FT} \left[ I(\mathbf{r}_\mathrm{0})\: e^{-i\frac{\mathbf{r}_\mathrm{0}^2 k \Delta z}{2f^2}} \right]_{\frac{k\Delta\mathbf{r}_{\perp}}{f}}\right\rvert}^2}{{\left\lvert\displaystyle\int{\mathrm{d}\mathbf{r}_\mathrm{0} \: I(\mathbf{r}_\mathrm{0})}\right\rvert}^2} \: .
\label{eq:mono_corr}
\end{equation}

\subsection{Bichromatic correlation function of the speckle}
\label{A4}
We consider now bichromatic correlation function $c_{2\lambda}(\mathbf{r}_{\perp},z,\lambda_\mathrm{p}, \lambda_\mathrm{c})$ between two speckle fields created at different wavelength speckle pattern \textit{at a single point} located by $\mathbf{r}=\lbrace{\mathbf{r}_\perp,z \rbrace}$ compared to the center of the Fourier plane. As discussed in the text, the correlation between the two speckles disorder potentials is directly linked to this function, see Eq.~\eqref{eq:bichromatic correlation}. It is defined as
\begin{equation}
c_{2\lambda}(\mathbf{r}_{\perp},z,\lambda_\mathrm{p}, \lambda_\mathrm{c})=\frac{\overline{\delta I (\mathbf{r}_{\perp},z,\lambda_\mathrm{p}) \delta I (\mathbf{r}_{\perp},z,\lambda_\mathrm{c})}}{\overline{I(\mathbf{r}_{\perp},z,\lambda_\mathrm{p})}\:\overline{I(\mathbf{r}_{\perp},z,\lambda_\mathrm{c})}} \: .
\end{equation}
The correlation function in the numerator is computed using Wick's theorem $\overline{\delta I (\mathbf{r}_{\perp},z,\lambda_\mathrm{p}) \delta I (\mathbf{r}_{\perp},z,\lambda_\mathrm{c})}= \left\lvert\Gamma_\mathcal{A}(\mathbf{r}_\perp, z, \lambda_\mathrm{p}, \lambda_\mathrm{c}) \right\rvert^2 $, with:
\begin{align}
&\left\lvert\Gamma_\mathcal{A}(\mathbf{r}_\perp, z, \lambda_\mathrm{p}, \lambda_\mathrm{c}) \right\rvert^2 = \left\lvert \overline{\mathcal{A}(\mathbf{r}_{\perp},z,\lambda_\mathrm{p}) \mathcal{A}^*(\mathbf{r}_{\perp},z,\lambda_\mathrm{c})} \right\rvert^2 \nonumber \\
& \nonumber= \left\lvert \frac{1}{\lambda_\mathrm{p}\lambda_\mathrm{c} z^2} \displaystyle\int{\mathrm{d}\mathbf{r}_\mathrm{0} \: I(\mathbf{r}_\mathrm{0}) \: e^{-i\frac{\mathbf{r}_\mathrm{0}^2 z}{2f^2}(k_\mathrm{p}-k_\mathrm{c})} \: }\right. \\
&\nonumber \times  e^{i \frac{\mathbf{r}_{\perp}.\mathbf{r}_\mathrm{0}}{f}(k_\mathrm{c}-k_\mathrm{p})} \displaystyle\int{\mathrm{d}\Delta\mathbf{r}_0 \: C_\mathrm{diff}(\Delta\mathbf{r}_0, \lambda_\mathrm{p}, \lambda_\mathrm{c})} \\
& \times \left. e^{i \frac{\Delta\mathbf{r}_0.\mathbf{r}_\mathrm{0} z}{f^2}\frac{k_\mathrm{p}+k_\mathrm{c}}{2}} \: e^{i\frac{\Delta\mathbf{r}_0.\mathbf{r}_{\perp}}{f}\frac{k_\mathrm{p}+k_\mathrm{c}}{2}}\right\rvert^2 \: .
\end{align}
Using (i) the same approximation as in~Eq.~\eqref{eq:cal_mono_corr} for the phase exponentials of the second integral, (ii) the normalization by the average intensity profile around the center $\lvert\overline{I(\mathbf{0},0)}\rvert^2$, and (iii) $\delta\lambda  = \lvert \lambda_\mathrm{c} - \lambda_\mathrm{p} \rvert \ll \lambda_\mathrm{p,c}\sim \lambda$, we obtain:
\begin{align}
&c_{2\lambda}(\mathbf{r}_{\perp},z,\lambda_\mathrm{p},\lambda_\mathrm{c}) \nonumber \\
&= \frac{\left\lvert \displaystyle\int{\mathrm{d}\Delta\mathbf{r}_0 \: C_\mathrm{diff}(\Delta\mathbf{r}_0,\lambda_\mathrm{p},\lambda_\mathrm{c})}\right\rvert^2}{\displaystyle\int{\mathrm{d}\Delta\mathbf{r}_0 \: C_\mathrm{diff}(\Delta\mathbf{r}_0,\lambda_\mathrm{p})}\displaystyle\int{\mathrm{d}\Delta\mathbf{r}_0 \: C_\mathrm{diff}(\Delta\mathbf{r}_0,\lambda_\mathrm{c})}} \nonumber \\
& \times \frac{\left\lvert\displaystyle\int{\mathrm{d}\mathbf{r}_\mathrm{0} \: I(\mathbf{r}_\mathrm{0}) \: e^{-ik \frac{\mathbf{r}_\mathrm{0}^2 z}{2f^2} \frac{\delta\lambda}{\lambda}} \: e^{ik\frac{\mathbf{r}_{\perp}.\mathbf{r}_\mathrm{0}}{f} \frac{\delta\lambda}{\lambda}}} \right\rvert^2}{\left\lvert\displaystyle\int{\mathrm{d}\mathbf{r}_\mathrm{0} \: I(\mathbf{r}_\mathrm{0})} \right\rvert^2} 
\end{align}
and consists of two terms. The first one only features the diffuser while the second describes the free space propagation after the diffuser. The first term can be computed using:
\begin{equation}
\int{\mathrm{d}\Delta\mathbf{r}_0 \: C_\mathrm{diff}(\Delta\mathbf{r}_0 ,\lambda)}= 2\pi r_\mathrm{diff}^2
\end{equation}

The normalized bichromatic correlation function finally reads as:
\begin{align}
&c_{2\lambda}(\mathbf{r}_{\perp},z,\lambda_\mathrm{p},\lambda_\mathrm{c}) \nonumber\\
&= C_\mathrm{diff}^2(\mathbf{0},\lambda_\mathrm{p},\lambda_\mathrm{c}) \frac{\left\lvert \mathrm{FT} \left[ I(\mathbf{r}_\mathrm{0}) \: e^{-ik\frac{\mathbf{r}_\mathrm{0}^2 z}{2f^2}}\right]_{\frac{k\mathbf{r}_{\perp}}{f} \frac{\delta\lambda}{\lambda}}\right\rvert^2}{\left\lvert\displaystyle\int{\mathrm{d}\mathbf{r}_\mathrm{0} \: I(\mathbf{r}_\mathrm{0})} \right\rvert^2} \nonumber\\
&= e^{-\sigma_{\Delta\phi}^2}\: c_{\mathrm{3D}}\left(\mathbf{r}_{\perp} \frac{\delta\lambda}{\lambda}, z \frac{\delta\lambda}{\lambda}\right)
\end{align}

%\bibliography{biblio_clean_Epjd.bib}

%

\end{document}